\documentclass[12pt]{iopart}
\usepackage{graphicx}
\usepackage{amssymb}
\usepackage{textcomp}
\usepackage{iopams}
\usepackage{psfrag}
\usepackage[nooneline,small,sc]{caption}
\begin{document}

\newenvironment{narrow}[2]{%
\begin{list}{}{%
\setlength{\topsep}{0pt}%
\setlength{\leftmargin}{#1}%
\setlength{\rightmargin}{#2}%
\setlength{\listparindent}{\parindent}%
\setlength{\itemindent}{\parindent}%
\setlength{\parsep}{\parskip}}%
\item[]}{\end{list}}

\title[Blue-detuned evanescent field surface traps for neutral atoms...]
{Blue-detuned evanescent field surface traps for neutral atoms
based on mode interference in ultra-thin optical fibres}
\author{G~Sagu\'{e}, A~Baade and A~Rauschenbeutel}
\address{Institut f\"ur Physik,
Universit\"at Mainz, 55128 
 Mainz, Germany}
\eads{\mailto{arno.rauschenbeutel@uni-mainz.de}}

\begin{abstract}
We present and analyze a novel concept for blue-detuned evanescent
field surface traps for cold neutral atoms based on two-mode
interference in ultra-thin optical fibres. When two or more
transverse modes with the same frequency co-propagate in the
fibre, their different phase velocities cause a stationary
interference pattern to establish. Intensity minima of the
evanescent field at any distance from the fibre surface can be
created and an array of optical microtraps can thus be obtained in
the evanescent field. We discuss three possible combinations of
the lowest order modes, yielding traps at one to two hundred
nanometres from the fibre surface which, using a few ten
milliwatts of trapping laser power, have a depth on the order of
1~mK for caesium atoms and a trapping lifetime exceeding
100~seconds. The resulting trapping geometry is of particular
interest because atoms in such microtrap arrays will be coupled to
any additional field propagating in the fibre via the evanescent
field, thereby realising ensembles of fibre-coupled atoms.
\end{abstract}
 \submitto{\NJP}

\section*{Introduction}

Recently, the production of ultra-thin optical fibres with
diameters smaller than the wavelength of the guided light has
become possible in a number of laboratories~\cite{Tong03, NicCh06,
Nayak07, Warken08}. Such fibres have attracted considerable
interest in the field of quantum optics due to their high
potential for efficiently coupling light and
matter~\cite{Balykin06, Nayak07, Sague07}. The guided modes in
such ultra-thin optical fibres exhibit a unique combination of
strong transverse confinement and pronounced evanescent
field~\cite{Balykin04}. Furthermore, the strong radial confinement
is maintained over the full length of the fibre waist, exceeding
the Rayleigh range of a comparably focused freely propagating
Gaussian beam by several orders of magnitude. This has been used
in a number of experiments to couple atoms and molecules to the
fibre mode via the evanescent field, showing that tapered optical
fibres (TOFs) are a powerful tool for their detection,
investigation, and manipulation: Recently, evanescent field
spectroscopy of a very small number of cold caesium atoms around a
500-nm diameter TOF has been performed~\cite{Sague07}. In a
similar experiment the fluorescence of resonantly irradiated atoms
around a 400-nm diameter TOF, coupled into the fibre mode, has
been detected and spectrally analyzed \cite{Nayak07,Nayak08}. The
absorbance of organic dye molecules, deposited on a
subwavelength-diameter TOF, has also been spectroscopically
characterized via the fibre transmission with unprecedented
sensitivity \cite{Warken07}. Finally, it has also been proposed to
trap atoms around ultra-thin fibres using the optical dipole force
exerted by the evanescent field \cite{Dowling96, Balykin04_2}. In
this case, the atoms can be coupled to and trapped near a
dielectric nanostructure
without the need of additional external light fields.\\

Here, we present a novel type of blue-detuned evanescent field
trap for cold neutral atoms based on two-mode interference in such
ultra-thin optical fibres. We consider a field-fibre configuration
where only the four lowest order modes propagate, the fundamental
mode, HE$_{11}$, and the first three higher order modes,
TE$_{01}$, TM$_{01}$ and HE$_{21}$. If the modes are coherently
excited, they will yield a stationary interference pattern while
co-propagating in the fibre due to their different phase
velocities. Previously, a similar scheme has been proposed for
trapping atoms in the evanescent field of a two-dimensional planar
waveguide. In this case, however, the interference of at least
four waveguide modes is required in order to achieve three
dimensional confinement of the atoms~\cite{Christandl04}.\\

We explore all possible pairs of modes that can be used to trap
cold neutral atoms in the intensity minima formed at the positions
of destructive interference in the evanescent field surrounding
the fibre. These combinations are HE$_{11}$+TE$_{01}$,
HE$_{11}$+HE$_{21}$ and TE$_{01}$+HE$_{21}$.\footnote{The
TM$_{01}$ mode cannot be used to create a trap because its
polarisation at the position of destructive interference cannot be
matched with any of the other modes due to its large component in
the direction of propagation.} We assume a cylindrical ultra-thin
silica fibre with 400 nm radius for all three trapping
configurations. The wavelength and the total power of the guided
light as well as the power distribution between the modes have
been chosen to fulfill the following criteria for each individual
trapping configuration: A three-dimensional trapping potential for
caesium atoms, a depth of the trap on the order of 1~mK, and a
trapping lifetime exceeding 100~seconds for an atom with an
initial
kinetic energy corresponding to a temperature of 100~$\mu$K.\\

This paper is organised as follows: Section~\ref{Section1} is
devoted to the analysis of the modal dispersion in ultra-thin
optical fibres and to the presentation of the electric field
equations in the evanescent field for the three considered modes.
In Sect.~\ref{HE11+TE01 trap} the trap arising from the
interference between the HE$_{11}$ and the TE$_{01}$ mode is
presented (HE$_{11}$+TE$_{01}$). Sections~\ref{HE11+HE21 trap}
and~\ref{HE21+TE01 trap} then treat the HE$_{11}$+HE$_{21}$ and
TE$_{01}$+HE$_{21}$ traps, respectively.\\

\section{Electric field and mode propagation in ultrathin optical
fibres}\label{Section1}

\subsection{Mode propagation}

We consider a step-index optical fibre consisting of a cylindrical
bulk of dielectric material with radius $a$ and refractive index
$n_{1}$, surrounded by a second dielectric medium with infinite
radius and refractive index $n_{2}$. For the guided light in such
a fibre a discrete set of propagation modes exists whose axial
propagation constant $\beta$ is fixed by the boundary
conditions~\cite{Yariv}. The number of modes and their axial
propagation constant is then determined by the radius of the fibre
$a$, the refractive indices of the two media, $n_{1}$ and $n_{2}$,
and the wavelength of the light $\lambda$ via the parameter
$V=(2\pi a/\lambda)\sqrt{n_{1}^2-n_{2}^2}$.\\

Figure~\ref{fig:Mode_betas_Paper} shows the axial propagation
constant $\beta$ for the first seven modes in the fibre normalized
to the wavenumber in free space $k_{0}$ as a function of the $V$
parameter. Note that $\beta/k_{0}$ lies between $n_{2}$ and
$n_{1}$ which is a condition that must be fulfilled by any
lossless mode~\cite{Yariv}. In the following, $\beta$ will be
referred to as propagation constant or phase velocity. The dashed
vertical line located at $V=3.11$ corresponds to the three
configurations considered in this paper: An ultra-thin optical
fibre of pure silica ($n_{1}=1,452$) with a radius $a=400$ nm,
surrounded by vacuum ($n_{2}=1$), and three similar wavelengths
$\lambda=849.0$ nm, 850.5 nm and 851.0 nm. In this case, only four
modes are allowed to propagate, the fundamental mode HE$_{11}$ and
the first three non-fundamental modes, i.e., TE$_{01}$, TM$_{01}$,
and HE$_{21}$. At this value of $V$ the phase velocities of all
modes differ significantly. This difference will cause an
interference pattern to establish along the fibre and, in
addition, results in different radial decay lengths of the
evanescent field outside the fibre for each mode. The modal
dispersion can therefore be used to create a tailored evanescent
field resulting from the interference of two or more
co-propagating modes.\\

\subsection{The fundamental HE$_{11}$ mode with quasi-linear
polarisation}

The $\vec{E}$ field equations of the fundamental HE$_{11}$ mode
with quasi-linear polarisation outside the
fibre, i.e., for $r>a$ are given by~\cite{Balykin04}:\\
\begin{eqnarray}
E_{x}(r,\phi,z,t)&=A_{11}\frac{\beta_{11}}{2q_{11}}\frac{J_{1}(h_{11}a)}{K_{1}(q_{11}a)}[(1-s_{11})K_{0}(q_{11}r)\cos(\varphi_{0})+\nonumber\\
&+(1+s_{11})K_{2}(q_{11}r)\cos(2\phi-\varphi_{0})]\exp[i(\omega t -\beta_{11} z)]\label{eq:Exlinoutside}\\
E_{y}(r,\phi,z,t)&=A_{11}\frac{\beta_{11}}{2q_{11}}\frac{J_{1}(h_{11}a)}{K_{1}(q_{11}a)}[(1-s_{11})K_{0}(q_{11}r)\sin(\varphi_{0})+\nonumber\\
&+(1+s_{11})K_{2}(q_{11}r)\sin(2\phi-\varphi_{0})]\exp[i(\omega t -\beta_{11} z)]\label{eq:Eylinoutside}\\
E_{z}(r,\phi,z,t)&=iA_{11}\frac{J_{1}(h_{11}a)}{K_{1}(q_{11}a)}K_{1}(q_{11}r)\cos(\phi-\varphi_{0})\exp[i(\omega
t -\beta_{11} z)]\label{eq:Ezlinoutside}
\end{eqnarray}
where,
\begin{eqnarray}
s_{11}&=&\Big[\frac{1}{(h_{11}a)^2}+\frac{1}{(q_{11}a)^2}\Big]\Big
[\frac{J'_{1}(h_{11}a)}{h_{11}aJ_{1}(h_{11}a)}+\frac{K'_{1}(q_{11}a)}{q_{11}aK_{1}(q_{11}a)}\Big]^{-1}\label{eq:s11}\\
h_{11}&=&\sqrt{k^2_{0}n^2_{1}-\beta_{11}^2}\label{eq:h11}\\
q_{11}&=&\sqrt{\beta_{11}^2-k^2_{0}n^2_{2}}\label{eq:q11}
\end{eqnarray}
In the equations above, $J'(x)$ ($K'(x)$) designates $dJ(x)/dx$
($dK(x)/dx$), $a$ denotes the radius of the fibre, $\beta_{11}$
the propagation constant of the HE$_{11}$ mode, and the angle
$\varphi_{0}$ gives the polarisation direction of the transverse
electric field $\vec{E}_{\perp}=(E_{x},E_{y})$, with
$\varphi_{0}=0$ leading to $x$-polarisation  and
$\varphi_{0}=\pi/2$ to $y$-polarisation. $A_{11}$ is a
normalisation constant of the fields that links the total power to
the maximal field amplitude~\cite{Bures}. The quantity $q_{11}$ in
Eq.~(\ref{eq:q11}) is particularly relevant since it fixes the
scale of the decay length of the fields outside the fibre which
can be defined as $\Lambda_{11}=1/q_{11}$. The HE$_{11}$ is a
hybrid mode since it is neither TE (transversal electric) nor TM
(transversal magnetic) because the axial field components  $E_{z}$
and $H_{z}$ are not zero~\cite{Yariv}. The designation of
quasi-linear polarisation stems from the fact that $E_{z}$ has a
$\pi/2$ dephasing with respect to $\vec{E}_{\perp}$, which results
in elliptical polarisation
except where $E_{z}=0$.\\

Figure~\ref{fig:Field_plot_HE11_lin} shows a vectorial plot of the
electric field component transverse to the fibre axis
$\vec{E}_{\perp}=(E_{x},E_{y})$ at $t=0$ and $z=0$, with
$\varphi_{0}$ set to zero. Note that the equations of the electric
field inside the fibre used in this figure are not explicitly
given here but can be found in~\cite{Balykin04}. The surface of
the fibre is indicated by a gray circle. The calculations have
been performed for a wavelength of $\lambda=850$~nm. The decay
length of the evanescent field for these parameters is
$\Lambda_{11}=164$ nm.

\subsection{The TE$_{01}$ mode}\label{The TE01 mode}

We now present the $\vec{E}$ field equations of the TE$_{01}$ mode
for $r>a$~\cite{Yariv}. As can be seen from
Fig.~\ref{fig:Mode_betas_Paper}, the TE$_{01}$ has, like the
TM$_{01}$, a cutoff value of $V=2.405$. This is the lowest cutoff
value of any non-fundamental mode and thus sets the single mode
condition of a step-index optical fibre.\\
\begin{eqnarray}
E_{\phi}(r,\phi,z,t) &=&
\frac{\omega\mu}{q_{01}}\frac{J_{0}(h_{01}a)}{K_{0}(q_{01}a)}B_{01}K_{1}(q_{01}r)\exp[i(\omega
t -\beta_{01} z)]\label{eq:TEoutside}\\
E_{z}(r,\phi,z,t)&=&E_{r}(r,\phi,z,t)= 0\nonumber
\end{eqnarray}
where,
\begin{eqnarray}
h_{01}&=&\sqrt{k^2_{0}n^2_{1}-\beta_{01}^2}\label{eq:h01}\\
q_{01}&=&\sqrt{\beta_{01}^2-k^2_{0}n^2_{2}}\label{eq:q01}
\end{eqnarray}
$\beta_{01}$ denotes the propagation constant of the TE$_{01}$
mode and $B_{01}$ is the normalisation constant of the field
amplitude. The TE$_{01}$ is classified as a transversal electric
mode since it has a vanishing $z$-component of the electric field.
Furthermore, the only non-vanishing electric field component is
$E_{\phi}$. The TE$_{01}$ thus possesses only one linearly
independent polarisation state. The corresponding orthogonal
polarisation state is the fibre eigenmode TM$_{01}$. These two
modes split due to the distinct influence of the fibre-vacuum
boundary on the different $\vec{E}$ polarisation
directions.\\

Figure~\ref{fig:Field_plot_TE01} shows a vectorial plot of the
$\vec{E}$ field at $t=0$ and $z=0$. The equations of the TE$_{01}$
mode inside the fibre used for this figure can be found in several
text books, see for example~\cite{Yariv,Snyder}. The calculations
have been performed for the same parameters as in
Fig.~\ref{fig:Field_plot_HE11_lin}. As shown in
Fig.~\ref{fig:Field_plot_TE01}, $\vec{E}$ vanishes at $r=0$,
which, together with the azimuthal symmetry of $E_{\phi}$,
produces a toroidal shape of the field amplitude distribution. The
decay length of the evanescent field for the given parameters is
$\Lambda_{01}=277$ nm.

\subsection{The HE$_{21}$ mode with quasi-linear polarisation}\label{The HE21 mode with quasi-linear polarization}

The field equations of the fundamental HE$_{21}$
mode with quasi-linear polarisation for $r>a$ are given by~\cite{Yariv}:\\
\begin{eqnarray}
E_{x}(r,\phi,z,t)&=-A_{21}\frac{\beta_{21}}{2q_{21}}\frac{J_{2}(h_{21}a)}{K_{2}(q_{21}a)}[(1-2s_{21})K_{1}(q_{21}r)\cos(\phi+2\phi_{0})+\nonumber\\
&+(1+2s_{21})K_{3}(q_{21}r)\cos(3\phi+2\phi_{0})]\exp[i(\omega t
-\beta_{21}
z)]\label{eq:HE21xlinoutside}\\
E_{y}(r,\phi,z,t)&=
A_{21}\frac{\beta_{21}}{2q_{21}}\frac{J_{2}(h_{21}a)}{K_{2}(q_{21}a)}[(1-2s_{21})K_{1}(q_{21}r)\sin(\phi+2\phi_{0})-\nonumber\\
&-(1+2s_{21})K_{3}(q_{21}r)\sin(3\phi+2\phi_{0})]\exp[i(\omega t
-\beta_{21}
z)]\label{eq:HE21ylinoutside}\\[4pt]
E_{z}(r,\phi,z,t)&=-iA_{21}\frac{J_{2}(h_{21}a)}{K_{2}(q_{21}a)}K_{2}(q_{21}r)\cos(2(\phi+\phi_{0}))\exp[i(\omega t -\beta_{21} z)]\nonumber\\
\label{eq:HE21zlinoutside}
\end{eqnarray}

where,
\begin{eqnarray}
s_{21}&=&\Big[\frac{1}{(h_{21}a)^2}+\frac{1}{(q_{21}a)^2}\Big]
\Big[\frac{J'_{2}(h_{21}a)}{h_{21}aJ_{2}(h_{21}a)}+\frac{K'_{2}(q_{21}a)}{q_{21}aK_{2}(q_{21}a)}\Big]^{-1}\label{eq:s21}\\
h_{21}&=&\sqrt{k^2_{0}n^2_{1}-\beta_{21}^2}\label{eq:h21}\\
q_{21}&=&\sqrt{\beta_{21}^2-k^2_{0}n^2_{2}}\label{eq:q21}
\end{eqnarray}
$\beta_{21}$ denotes the propagation constant of the HE$_{21}$
mode and $A_{21}$ is the normalisation constant of the field
amplitude. $\phi_{0}$ determines the polarisation direction of
$\vec{E}_{\perp}$, with $\phi_{0}=0$ and $\phi_{0}=\pi/4$ leading
to two orthogonal polarisation states of the transverse electric
field. The HE$_{21}$ is a hybrid mode with six non-vanishing
components of the $\vec{E}$ and $\vec{H}$ fields. Like for the
HE$_{11}$, the designation of quasi-linear polarisation stems from
the fact that $E_{z}$ has a $\pi/2$ dephasing with respect to
$\vec{E}_{\perp}$.\\

\begin{figure}[hbtp]
\begin{narrow}{-1.0cm}{-0.3cm}
\begin{minipage}[t]{0.6\textwidth}
\psfrag{HE1}[][][0.9]{HE$_{11}$} \psfrag{TM01}[][][0.9]{TM$_{01}$}
\psfrag{TE01}[][][0.9]{TE$_{01}$}\psfrag{HE21}[][][0.9]{HE$_{21}$}
\psfrag{HE12}[][][0.9]{HE$_{12}$}
\psfrag{HE31}[][][0.9]{HE$_{31}$}\psfrag{EH1}[c][r][0.9]{EH$_{11}$}
\psfrag{beta}[][][0.9]{$\beta/k_{0}$}
\psfrag{V}[l][r][0.9]{$V=\frac{2\pi
a}{\lambda}\sqrt{n^2_{1}-n^2_{2}}$}\psfrag{n1}[][][0.8]{$n_{1}$}
\psfrag{n2}[][][0.8]{$n_{2}$}\psfrag{0}[][][0.8]{0}\psfrag{1}[][][0.8]{1}
\psfrag{2}[][][0.8]{2}\psfrag{3}[][][0.8]{3}
\psfrag{4}[][][0.8]{4}\psfrag{5}[][][0.8]{5}
\includegraphics[width=0.92\textwidth]{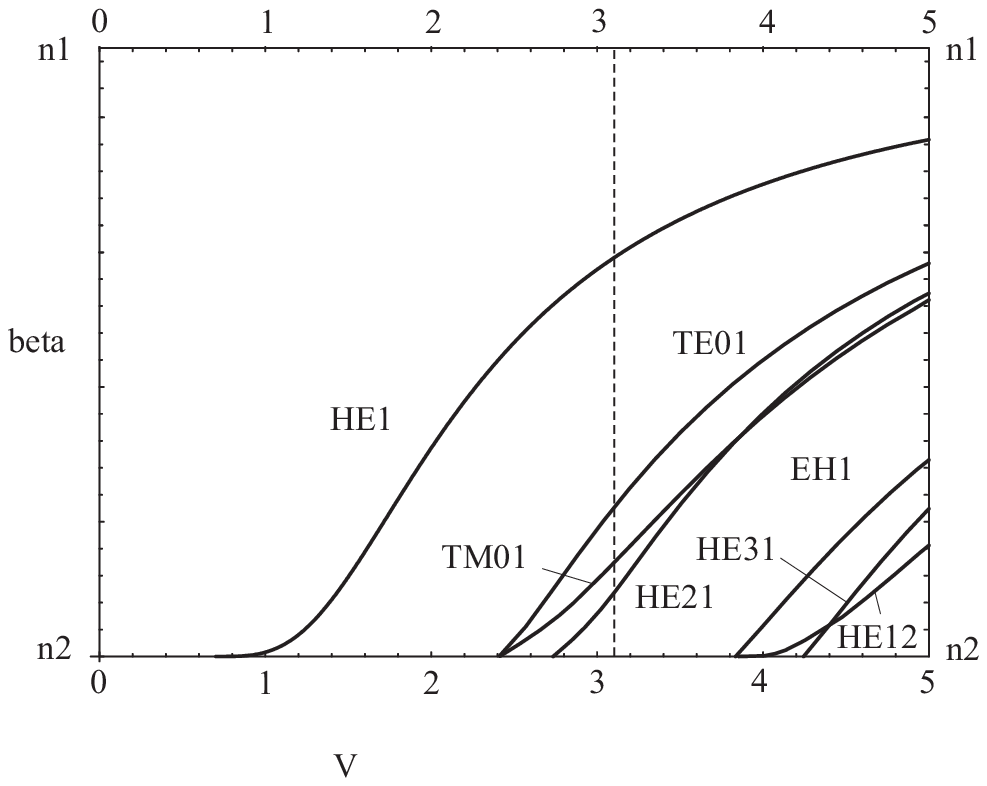}
\caption{Normalised propagation constant $\beta/k_{0}$ versus $V$
parameter for the first seven modes in the fibre. The dashed
vertical line is located at $V=3.11$ which corresponds to the
three trapping configurations considered in this paper.
\label{fig:Mode_betas_Paper}}
\end{minipage}%
\begin{minipage}[t]{0.6\textwidth}
\centering \psfrag{p100}[][][0.8]{0.4}\psfrag{x}[t][c][1]{$x$
($\mu$m)} \psfrag{y}[b][c][1]{$y$
($\mu$m)}\psfrag{p000}[][][0.8]{0}\psfrag{n100}[][][0.8]{-0.4}
\includegraphics[totalheight=7cm]{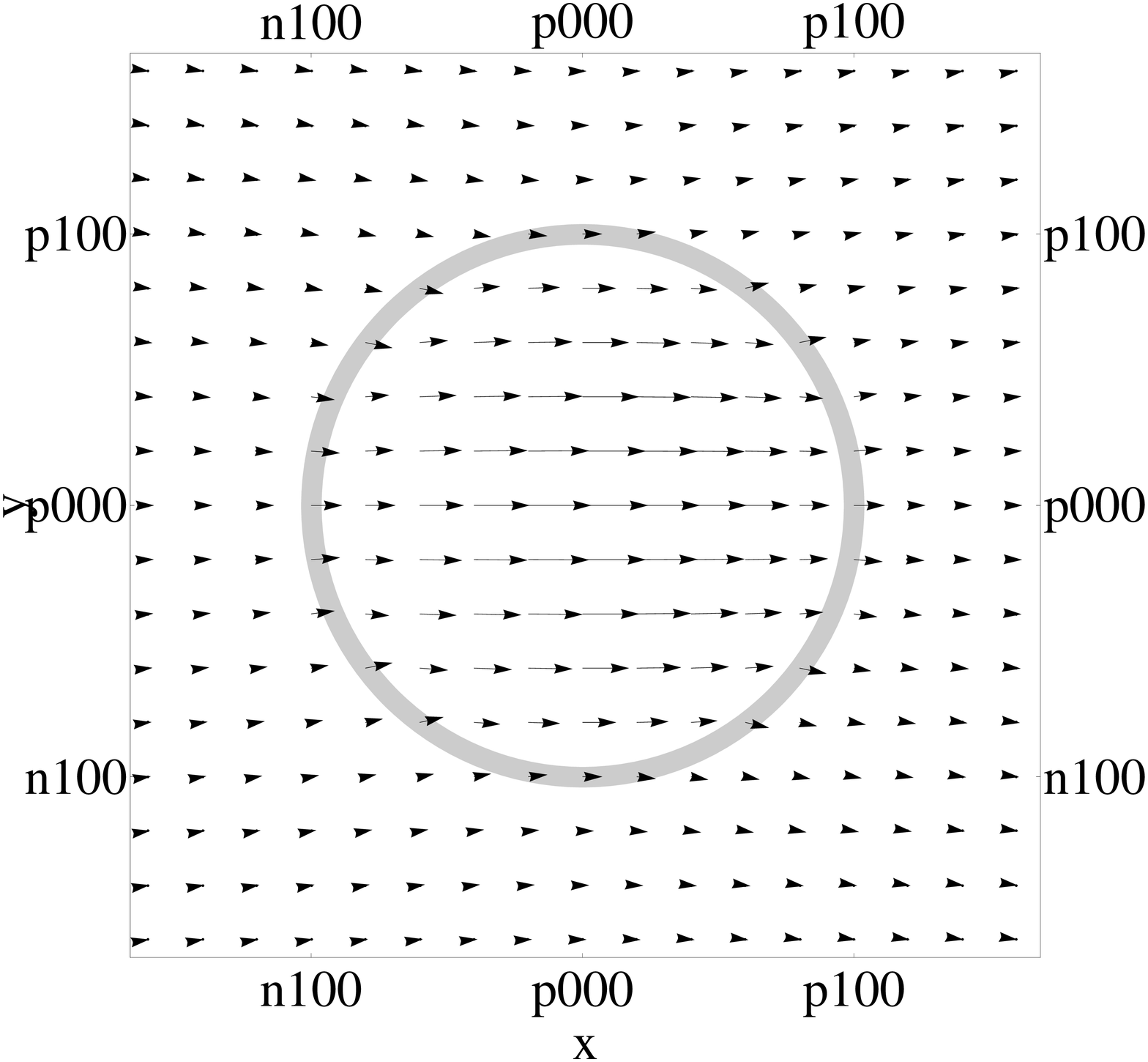}
\caption{Field plot of the electric field component perpendicular
to the fibre axis $\vec{E_{\perp}}=(E_{x},E_{y})$ for the
HE$_{11}$ mode at $t=0$, $z=0$ and for $\varphi_{0}=0$ (see
Eqs.~(\ref{eq:Exlinoutside}) and~(\ref{eq:Eylinoutside})). The
fibre is indicated by the gray circle. The following parameters
have been used: $a=400$~nm, $n_{1}=1.452$, $n_{2}=1$, and
$\lambda=850$~nm.\label{fig:Field_plot_HE11_lin}}
\end{minipage}\\[20pt]
\begin{minipage}[t]{0.6\textwidth}
\centering \psfrag{p100}[][][0.8]{0.4}\psfrag{x}[t][c][1]{$x$
($\mu$m)}\psfrag{y}[b][c][1]{$y$
($\mu$m)}\psfrag{p000}[][][0.8]{0}\psfrag{n100}[][][0.8]{-0.4}
\includegraphics[totalheight=7cm]{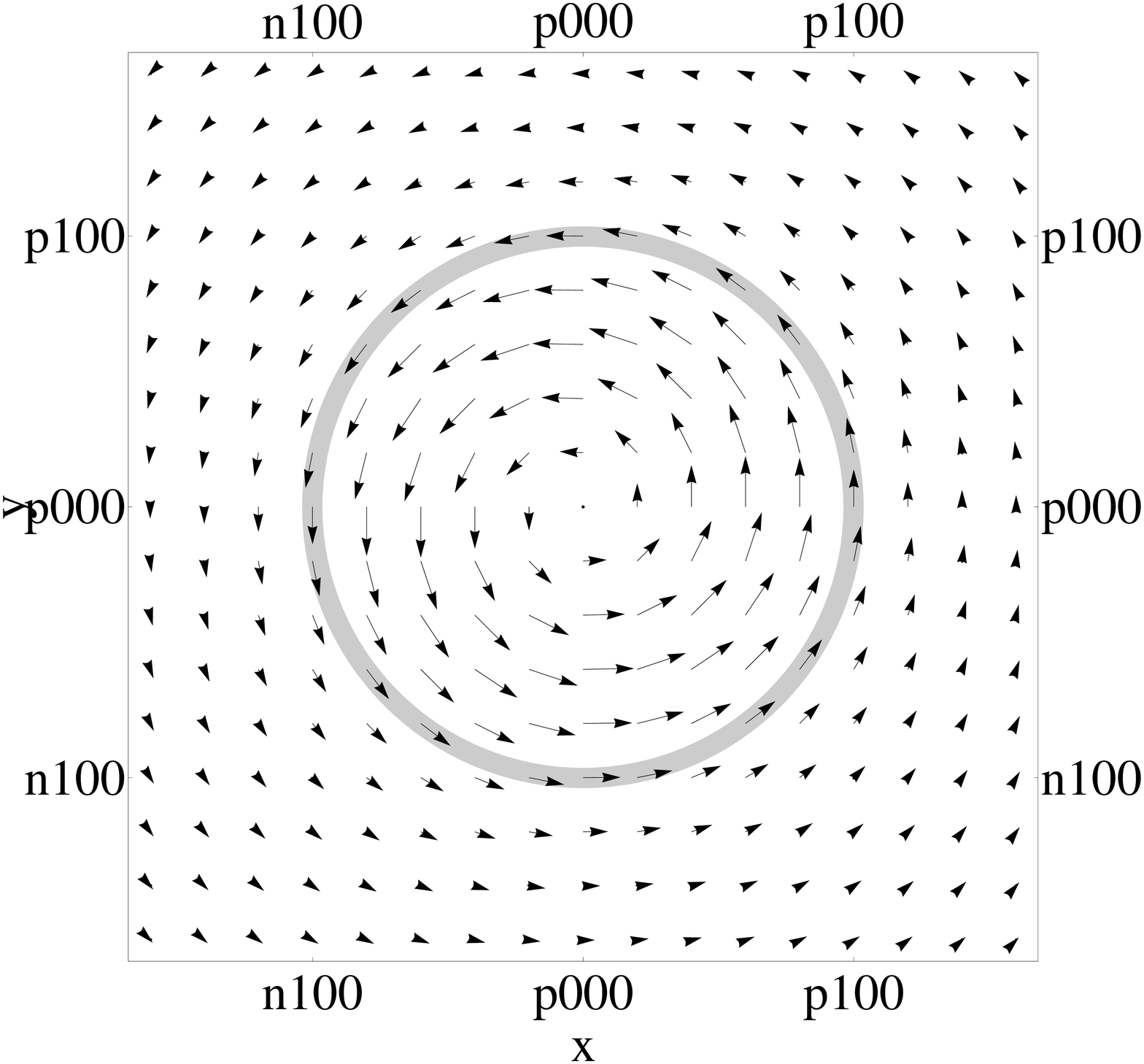}
\caption{Field plot of the electric field $\vec{E}$ for the
TE$_{01}$ mode at $t=0$ and $z=0$ (see Eq.~(\ref{eq:TEoutside})).
The fibre is indicated by the gray circle. The fibre parameters
are identical to
Fig.~\ref{fig:Field_plot_HE11_lin}.\label{fig:Field_plot_TE01}}
\end{minipage}%
\begin{minipage}[t]{0.6\textwidth}
\centering \psfrag{p100}[][][0.8]{0.4}\psfrag{x}[t][c][1]{$x$
($\mu$m)}\psfrag{y}[b][c][1]{$y$
($\mu$m)}\psfrag{p000}[][][0.8]{0}\psfrag{n100}[][][0.8]{-0.4}
\includegraphics[totalheight=7cm]{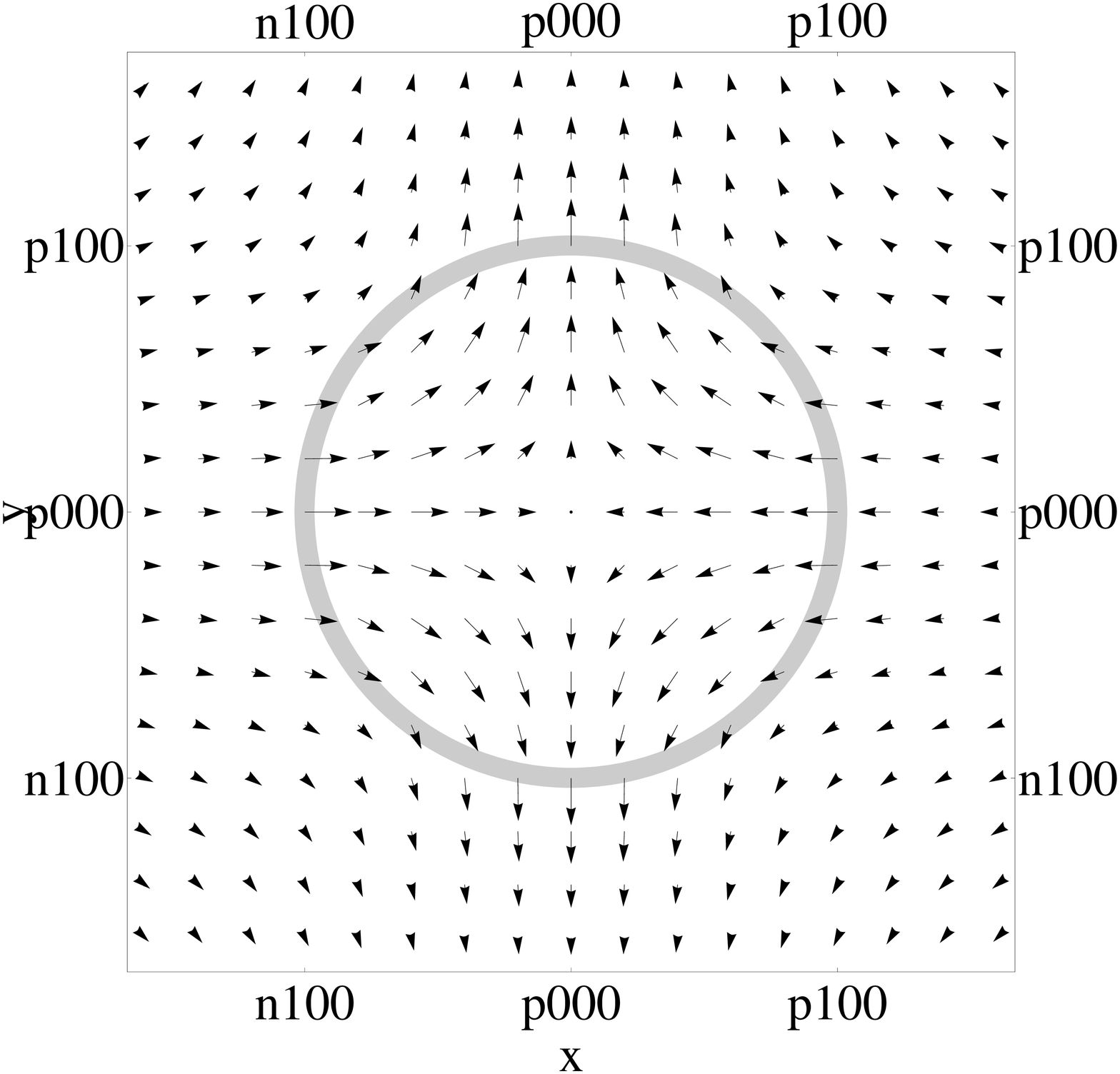}
\caption{Field plot of the electric field component perpendicular
to the fibre axis $\vec{E_{\perp}}=(E_{x},E_{y})$ for the
HE$_{21}$ mode at $t=0$, $z=0$ and for $\phi_{0}=0$ (see
Eqs.~(\ref{eq:HE21xlinoutside}) and~(\ref{eq:HE21ylinoutside})).
The fibre is indicated by the gray circle. The fibre parameters
are identical to
Fig.~\ref{fig:Field_plot_HE11_lin}.\label{fig:Field_plot_HE21_linear}}
\end{minipage}
\end{narrow}
\end{figure}

Figure~\ref{fig:Field_plot_HE21_linear} shows a vectorial plot of
the electric field component transversal to the fibre axis
$\vec{E}_{\perp}=(E_{x},E_{y})$ at $t=0$ and $z=0$ with $\phi_{0}$
set to zero. Again, the equations of the HE$_{21}$ mode inside the
fibre used for this figure can be found in several text
books~\cite{Yariv,Snyder}. The calculations have been performed
for the same parameters as in Fig.~\ref{fig:Field_plot_HE11_lin}
with a polarisation direction given by $\phi_{0}=0$. The decay
length of the evanescent field for these parameters is
$\Lambda_{21}=420$ nm.\\

\section{HE$_{\textbf{{\scriptsize 11}}}$+TE$_{\textbf{{\scriptsize 01}}}$ trap}\label{HE11+TE01 trap}

We now show that an evanescent field surface trap for cold atoms
can be obtained from the interference between the fundamental
HE$_{11}$ mode, which we assume to be quasi-linearly polarised,
and the TE$_{01}$
mode. 
By choosing the appropriate power distribution between the modes,
an array of local minima of the field intensity at any distance
from the fibre surface can be created at the positions where the
two fields optimally cancel. For blue-detuned light with respect
to the atomic transition frequency a dipole force proportional to
the negative gradient of the field intensity is then exerted on
the atoms~\cite{Balykin04_2} confining them in
the intensity minima.\\

As an example, we discuss the properties of the above trap for
caesium atoms. The trap can be created with 50~mW of light at a
wavelength of 850.5~nm and the same fibre parameters as in
Sect.~\ref{Section1}. This power is realistic for such a fibre in
vacuum: We could experimentally show that appropriately produced
fibres with an even smaller radius of 250~nm carry more than
300~mW of power in such conditions without fusing. With 72\% of
the power propagating in the HE$_ {11}$ mode and 28\% in the
TE$_{01}$ mode, a trap for cold caesium atoms with a trapping
minimum at 134~nm from the fibre surface is formed. The depth of
the trap is 0.92~mK and the trapping lifetime resulting from
heating due to spontaneous scattering of photons exceeds 100
seconds for caesium atoms
with an initial kinetic energy corresponding to 100~$\mu$K.\\

\begin{figure}[hbtp]
\begin{narrow}{-2.3cm}{2.7cm}
\begin{minipage}[t]{0.66\textwidth}
\centering\psfrag{x}[t][c][1]{$x$ (nm)} \psfrag{y}[b][c][1]{$y$
(nm)} \psfrag{b1}[c][c][0.55]{0.2}\psfrag{b2}[c][t][0.55]{0.5}
\psfrag{b3}[c][r][0.55]{2.5}
\psfrag{b4}[l][rt][0.55]{10}\psfrag{b5}[c][rt][0.55]{50}\psfrag{b6}[b][t][0.55]{0.9}
\psfrag{b7}[c][t][0.55]{0.5}\psfrag{b8}[c][t][0.55]{0.2}\psfrag{b9}[c][r][0.55]{0.9}
\psfrag{0}[][][0.8]{0}\psfrag{n500}[][][0.8]{-500}\psfrag{n1000}[][][0.8]{-1000}
\psfrag{500}[][][0.8]{500}\psfrag{1000}[][][0.8]{1000}
\includegraphics[totalheight=8.5cm]{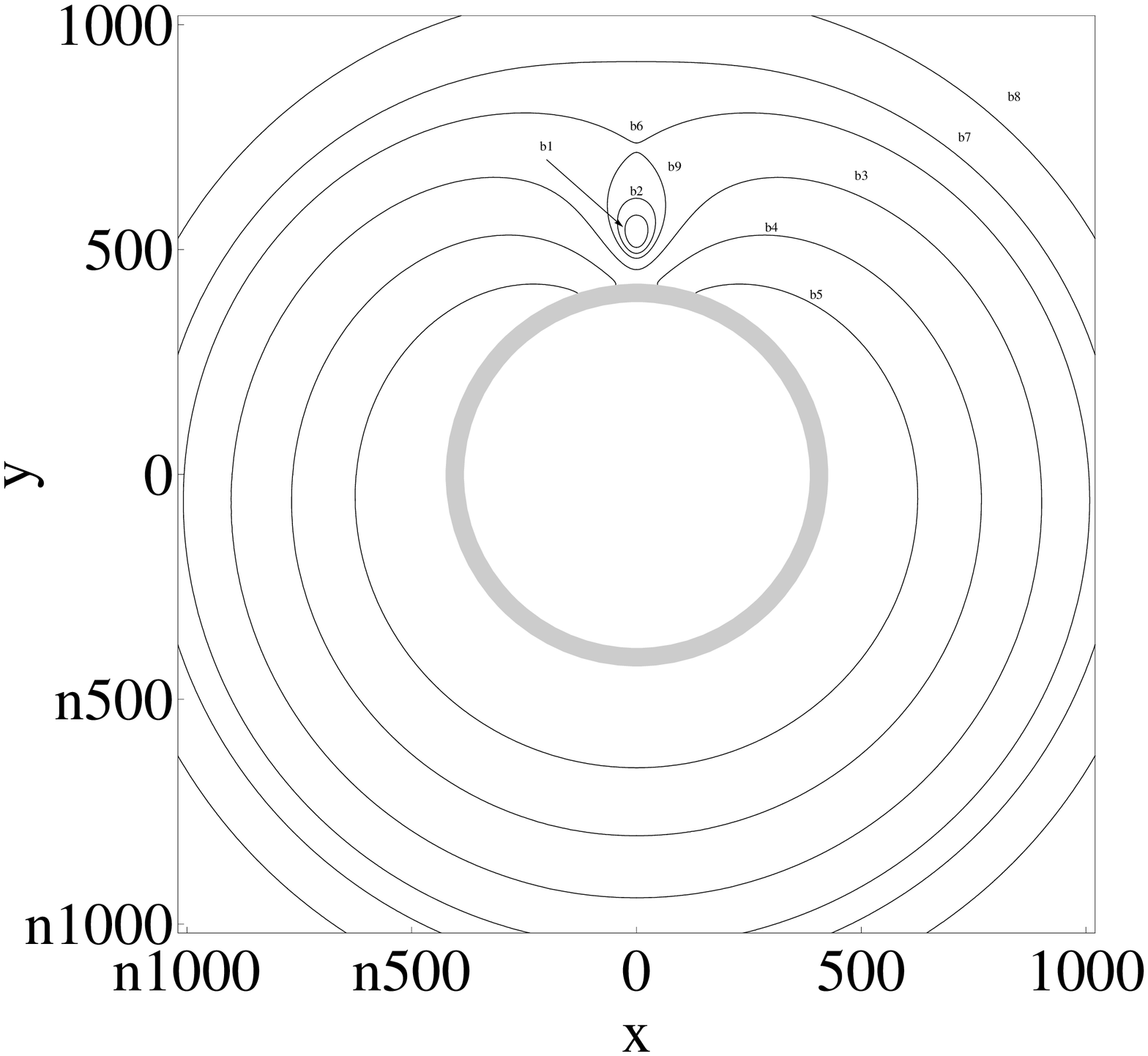}
\caption{Contour plot of the HE$_{11}$+TE$_{01}$ trap in the plane
$z=4.61$ $\mu$m for the following parameters: $P=50$~mW,
$\tau=0.72$, $\lambda=850.5$~nm, $a=400$~nm, $n_{1}=1.452$, and
$n_{2}=1$. The fibre surface is indicated by the gray circle and
the equipotential lines are labelled in
mK.\label{fig:TE01_HE11_az_all}}
\end{minipage}%
\begin{minipage}[t]{0.66\textwidth}
\centering\psfrag{x}[t][c][1]{$y$ ($\mu$m)}
\psfrag{z}[b][c][1]{$z$ ($\mu$m)}
\psfrag{b1}[c][c][0.55]{0.5}\psfrag{b2}[c][c][0.55]{0.9}\psfrag{b3}[][][0.55]{4}
\psfrag{b4}[c][lt][0.55]{25}\psfrag{b5}[c][c][0.55]{150}\psfrag{b6}[c][r][0.55]{0.9}
\psfrag{b7}[c][r][0.55]{0.5}
\psfrag{0}[][][0.8]{0}\psfrag{n1}[][][0.8]{-1}\psfrag{nm5}[][][0.8]{-0.5}
\psfrag{0.5}[][][0.8]{0.5}\psfrag{3}[][][0.8]{3}\psfrag{4}[][][0.8]{4}\psfrag{5}[][][0.8]{5}
\psfrag{6}[][][0.8]{6}\psfrag{7}[][][0.8]{7}\psfrag{1}[][][0.8]{1}
\includegraphics[totalheight=8.5cm]{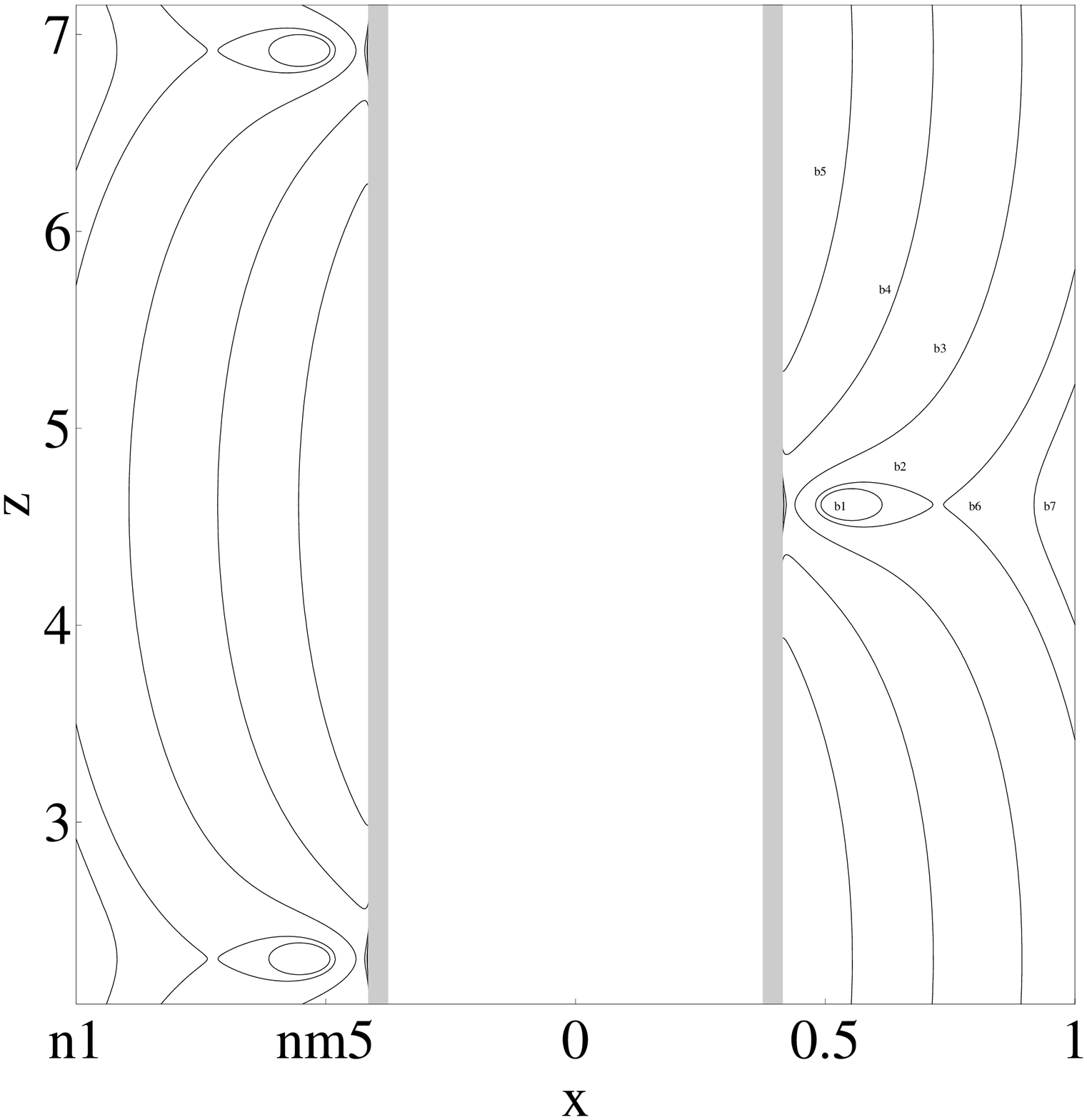}
\caption{Contour plot of the HE$_{11}$+TE$_{01}$ trap in the plane
$x=0$ for the same parameters as in
Fig.~\ref{fig:TE01_HE11_az_all}. The fibre surface is indicated by
the two vertical gray lines and the equipotential lines are
labelled in mK.\label{fig:TE01_HE11_long_all}}
\end{minipage}\\[20pt]
\end{narrow}

\centering\psfrag{n02}[][][0.8]{-0.2}\psfrag{0.0}[][][0.8]{0.0}\psfrag{0.2}[][][0.8]{0.2}
\psfrag{0.4}[][][0.8]{0.4}\psfrag{0.6}[][][0.8]{0.6}\psfrag{0.8}[][][0.8]{0.8}
\psfrag{1.0}[][][0.8]{1.0}\psfrag{1.2}[][][0.8]{1.2}\psfrag{1.4}[][][0.8]{1.4}
\psfrag{1.6}[][][0.8]{1.6}\psfrag{1.8}[][][0.8]{1.8}\psfrag{2.0}[][][0.8]{2.0}
\psfrag{2.2}[][][0.8]{2.2}
\psfrag{400}[][][0.8]{400}\psfrag{500}[][][0.8]{500}\psfrag{600}[][][0.8]{600}
\psfrag{700}[][][0.8]{700}\psfrag{800}[][][0.8]{800}\psfrag{1000}[][][0.8]{1000}
\psfrag{1200}[][][0.8]{1200} \psfrag{x}[][][1]{$y$
(nm)}\psfrag{Potential}[][][1]{potential energy (mK)}
\includegraphics[totalheight=8.5cm]{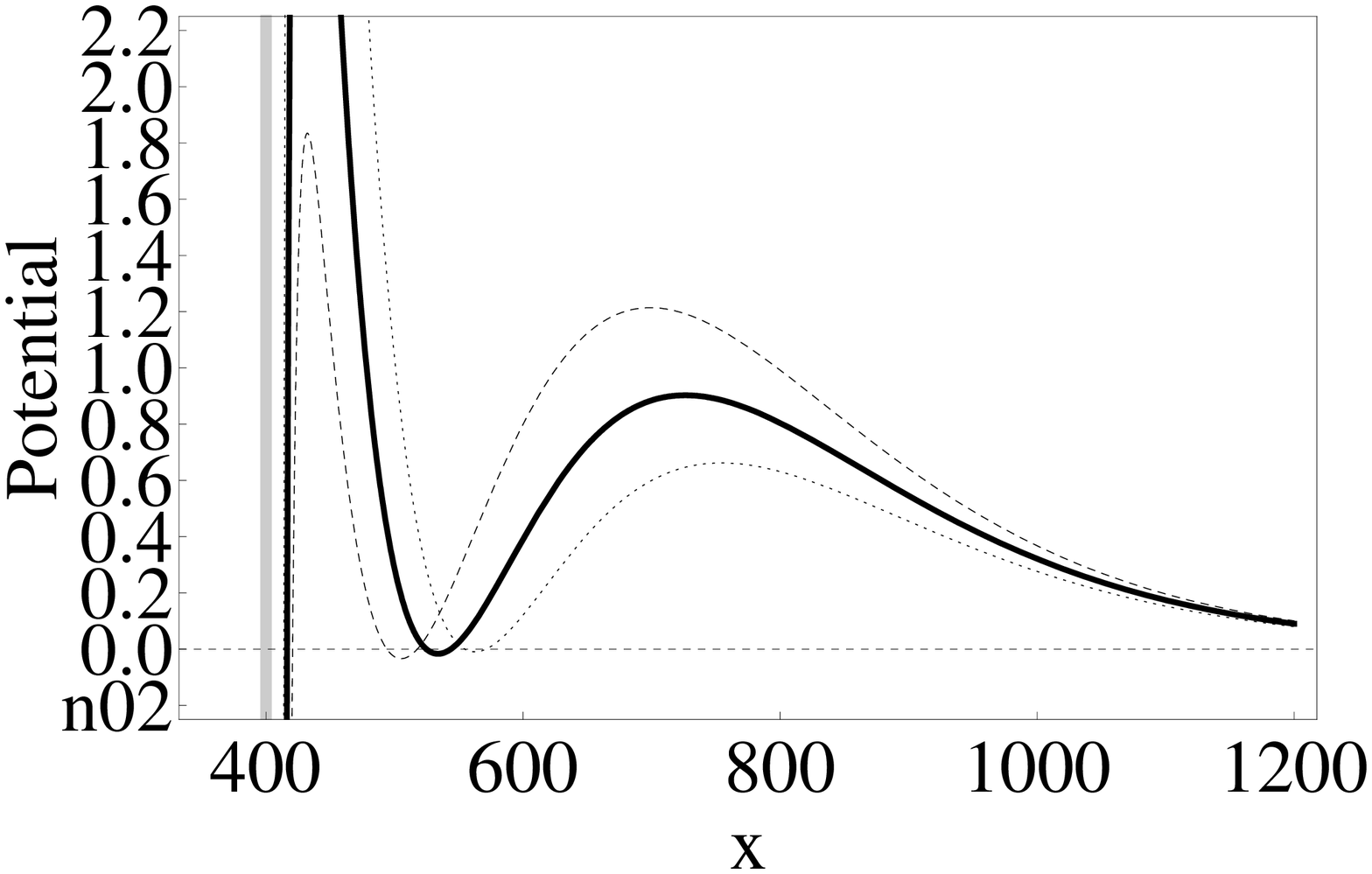}
\caption{Plot of the trapping potential versus the position along
the $y$ axis for P$_{11}=\tau_{0}$P (solid line),
P$_{11}=(\tau_{0}+\sigma)$P (dotted line) and
P$_{11}=(\tau_{0}-\sigma)$P (dashed line). The parameters are the
same as in Fig.~\ref{fig:TE01_HE11_az_all}. The fibre surface is
indicated by the vertical gray line.\label{fig:TE01_HE11_rad}}

\end{figure}

Figure~\ref{fig:TE01_HE11_az_all} shows a contour plot of the
trapping potential including the van der Waals surface
potential~\cite{Chevrollier} in the plane $z=4.61$~$\mu$m. For the
calculations, we use the van der Waals potential of an infinite
planar silica surface~\cite{Balykin04_2}. The fibre surface is
indicated by a gray circle and the equipotential lines are
labelled in mK. The trapping minimum is located at $\phi=\pi/2$,
$r=534$~nm and $z=4.61$~$\mu$m. The trapping minimum lies on the
$y$-axis because here the polarisation of the two modes matches
and the interference is maximally destructive. This polarisation
matching between the two modes can be understood when comparing
Figs.~\ref{fig:Field_plot_HE11_lin} and~\ref{fig:Field_plot_TE01}.
Note that while destructive interference takes place at
$\phi=\pi/2$, there is constructive interference at $\phi=3\pi/2$.
When varying the $\varphi_{0}$ parameter in
Eqs.~(\ref{eq:Exlinoutside}),~(\ref{eq:Eylinoutside})
and~(\ref{eq:Ezlinoutside}), the polarisation direction of the
HE$_{11}$ mode can be turned and thereby the azimuthal position of
the trap can be varied because the potential has a
$\cos(\phi-\varphi_{0})$ dependence. Using a harmonic potential
approximation, we calculate the azimuthal oscillation frequency to
be $\omega_{\phi}/2\pi\approx1.07$~MHz. The extension of the trap
volume in the azimuthal direction for caesium atoms with a kinetic
energy corresponding to 100~$\mu$K
is 34~nm.\\

Figure~\ref{fig:TE01_HE11_long_all} shows the contour plot of the
trapping potential in the plane $x=0$. The fibre surface is
indicated by two vertical gray lines. The interference between the
modes creates an array of traps in the axial direction with a
periodicity given by the beat length of the two co-propagating
modes, $z_{0}=2\pi/(\beta_{11}-\beta_{01})=4.61$~$\mu$m. In
addition, there is a second array of traps on the opposite side of
the fibre with same periodicity which is shifted by $z_{0}/2$. The
potential has a $\sin((\beta_{11}-\beta_{01})z)$ dependence in the
axial direction. The axial trapping frequency is calculated to be
$\omega_{z}/2\pi\approx528$~kHz. The extension of the trap volume
in this direction for caesium atoms with a kinetic energy
corresponding
to 100~$\mu$K is 68~nm.\\

Figure~\ref{fig:TE01_HE11_rad} shows the trapping potential along
the $y$-axis. The fibre surface is indicated by a vertical gray
line. The solid black line corresponds to the sum of the
light-induced potential and the van der Waals potential when 72\%
of the power propagates in the HE$_ {11}$ mode. The dashed and
dotted lines correspond to the same potential assuming slightly
different power distributions between the modes: We define the
parameter $\tau$ such that P$_{11}=\tau$P and P$_{01}=(1-\tau)$P,
where P denotes the total power transmitted through the fibre,
P$_{11}$ the power propagating in the HE$_{11}$ mode, and P$_{01}$
the power propagating in the TE$_{01}$ mode. We assume that $\tau$
can be controlled with a precision of
$\sigma=0.05\sqrt{\tau_{0}(1-\tau_{0})}$, i.e., $\sigma=0.025$ for
$\tau_{0}=0.5$. For the case of $\tau_{0}=0.72$ the power
distribution between the modes $\tau$ would then be controlled
within $\pm 0.022$. We consider this value to be a conservative
assumption for the precision of the power distribution between the
two modes. For the case of P$_{11}=(\tau_{0}+\sigma)$P (dotted
line) the trap is 27\% shallower compared to the trap for
P$_{11}=\tau_{0}$P (solid line), whereas for the case of
P$_{11}=(\tau_{0}-\sigma)$P (dashed line) the trap is 30\% deeper.
While the trap depth increases when decreasing $\tau$, the
trapping minimum is also shifted towards the fibre. When further
decreasing $\tau$ the depth of the trap thus drastically reduces
because the van der Waals potential becomes larger than the
light-induced potential. Furthermore, the potential barrier in the
direction towards the fibre becomes narrower which would
eventually lead to tunnelling of the atoms. The parameters
presented here have been chosen in such a way that even with
realistic experimental uncertainties the trap remains sufficiently
deep and the tunnelling is negligible compared to the trapping
lifetime. Note that the total potential is negative at its minimum
due to the influence of the van der Waals potential. Since the
$z$-component of the electric field in the HE$_{11}$ mode vanishes
at $\phi=\pi/2$, the polarisation in the two modes perfectly
matches at the intensity minimum and the van der Waals potential
at this position is the only influence on the atoms. We calculate
the radial trapping frequency to be
$\omega_{r}/2\pi\approx770$~kHz and the extension of the trapping
volume in the radial direction for caesium atoms with a kinetic
energy corresponding
to 100~$\mu$K is 47~nm.\\

The calculations of the lifetime have been performed assuming
caesium atoms with an initial kinetic energy equivalent to
100~$\mu$K trapped in a three dimensional classical harmonic
potential with oscillation amplitudes corresponding to the above
given extensions of the trapping volume. Note that the trap is not
perfectly symmetric along the y-axis (see
Fig.~\ref{fig:TE01_HE11_rad}), we account for this fact by biasing
the oscillation amplitude in this direction. From that, the mean
squared field amplitude at the position of the atom has been
calculated by integrating over all possible classical oscillation
modes. Using this method, we find a scattering rate of
39~photons/second and a trapping lifetime of 108~seconds.\\

\section{HE$_{\textbf{{\scriptsize 11}}}$+HE$_{\textbf{{\scriptsize 21}}}$ trap}\label{HE11+HE21 trap}

We now consider the trap arising from the interference between the
HE$_{11}$ and the HE$_{21}$ mode. It is created using 25~mW of
light at a wavelength of 849.0~nm and the same fibre parameters as
in Sect.~\ref{Section1}. The polarisation orientation of the modes
has been chosen such that the trap forms at $\phi=0$. This
corresponds to $\varphi_{0}=\phi_{0}=0$ in
Eqs.~(\ref{eq:Exlinoutside}) to~(\ref{eq:Ezlinoutside})
and~(\ref{eq:HE21xlinoutside}) to~(\ref{eq:HE21zlinoutside}),
respectively. With 84\% of the power propagating in the HE$_{11}$
mode, i.e., $\tau=0.84$ (see Sect.~\ref{HE11+TE01 trap}), a trap
at 152~nm from the fibre surface is formed. The depth of the trap
is 1.2~mK and the trapping lifetime resulting from spontaneous
scattering of photons exceeds 100~seconds for caesium atoms
with an initial kinetic energy corresponding to 100~$\mu$K.\\

\begin{figure}[hbtp]
\begin{narrow}{-2.2cm}{3cm}
\begin{minipage}[t]{0.65\textwidth}
\centering\psfrag{x}[t][c][1]{$x$ (nm)} \psfrag{y}[b][c][1]{$y$
(nm)}
\psfrag{b1}[c][rt][0.55]{0.5}\psfrag{b2}[][][0.55]{1.3}\psfrag{b3}[c][c][0.55]{1.6}
\psfrag{b4}[b][t][0.55]{1.3}\psfrag{b5}[c][c][0.55]{3}\psfrag{b6}[][][0.55]{10}
\psfrag{b7}[c][l][0.55]{1.6}\psfrag{b8}[c][lt][0.55]{0.5}\psfrag{b9}[c][l][0.55]{0.1}
\psfrag{0}[][][0.8]{0}\psfrag{n500}[][][0.8]{-500}\psfrag{n1000}[][][0.8]{-1000}
\psfrag{500}[][][0.8]{500}\psfrag{1000}[][][0.8]{1000}
\includegraphics[totalheight=8.5cm]{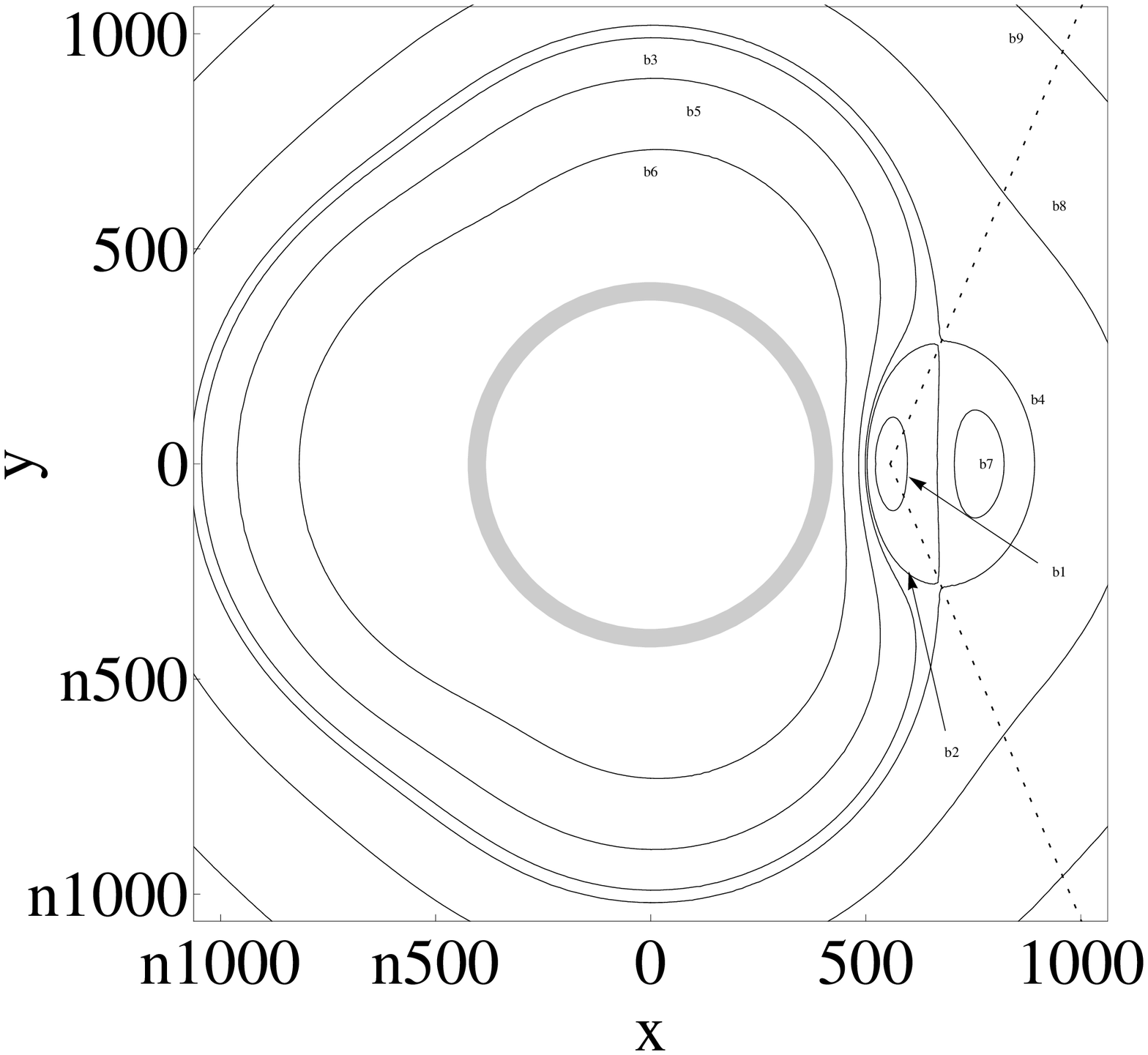}
\caption{Contour plot of the HE$_{11}$+HE$_{21}$ trap in the plane
$z=3.45$~$\mu$m for the following parameters: $P=25$~mW,
$\tau=0.84$, $\lambda=849.0$~nm, $a=400$~nm, $n_{1}=1.452$, and
$n_{2}=1$. The fibre surface is indicated by the gray circle and
the equipotential lines are labelled in
mK.\label{fig:HE21_HE11_az_all}}
\end{minipage}%
\begin{minipage}[t]{0.695\textwidth}
\centering\psfrag{x}[t][c][1]{$x$ ($\mu$m)}
\psfrag{z}[b][c][1]{$z$ ($\mu$m)}
\psfrag{b1}[c][tr][0.55]{0.6}\psfrag{b2}[l][r][0.55]{1.4}\psfrag{b3}[b][t][0.55]{5}
\psfrag{b4}[c][l][0.55]{20}\psfrag{b5}[rt][lb][0.6]{100}\psfrag{b6}[c][c][0.6]{1.4}
\psfrag{0.}[][][0.8]{0}\psfrag{n10}[][][0.8]{-1}\psfrag{nm5}[][][0.8]{-0.5}
\psfrag{0.5}[][][0.8]{0.5}\psfrag{1.}[][][0.8]{1}\psfrag{1}[][][0.8]{1}
\psfrag{2}[][][0.8]{2}\psfrag{3}[][][0.8]{3}\psfrag{4}[][][0.8]{4}\psfrag{5}[][][0.8]{5}

\includegraphics[totalheight=8.5cm]{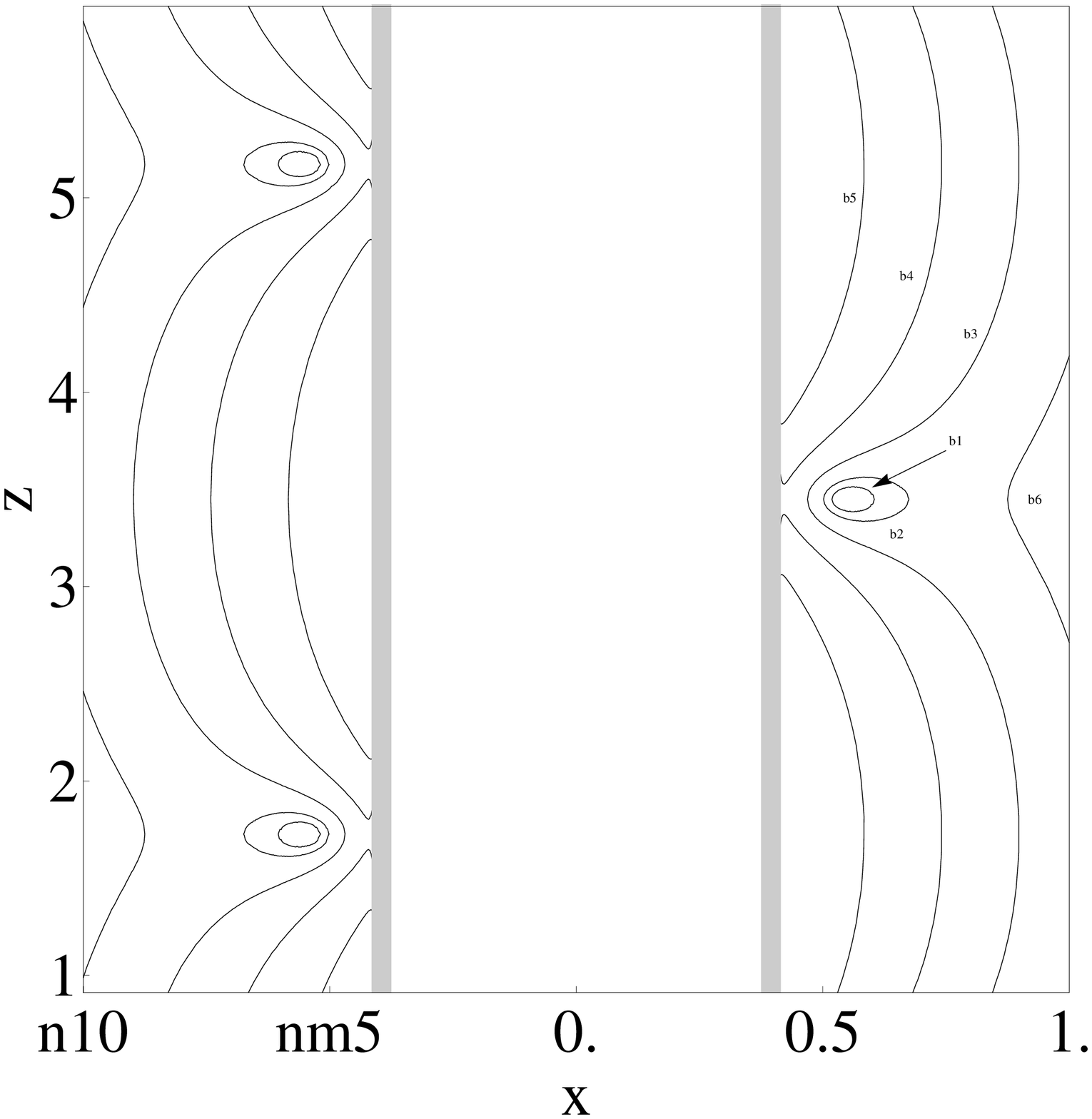}
\caption{Contour plot of the HE$_{11}$+HE$_{21}$ trap in the plane
$y=0$ for the same parameters as in
Fig.~\ref{fig:HE21_HE11_az_all}. The fibre surface is indicated by
the two vertical gray lines and the equipotential lines are
labelled in mK. \label{fig:HE21_HE11_long_all}}
\end{minipage}\\[20pt]
\begin{minipage}[t]{0.65\textwidth}
\centering\psfrag{n02}[][][0.8]{-0.2}\psfrag{0.0}[][][0.8]{0.0}\psfrag{0.2}[][][0.8]{0.2}
\psfrag{0.4}[][][0.8]{0.4}\psfrag{0.6}[][][0.8]{0.6}\psfrag{0.8}[][][0.8]{0.8}
\psfrag{1.0}[][][0.8]{1.0}\psfrag{1.2}[][][0.8]{1.2}\psfrag{1.4}[][][0.8]{1.4}
\psfrag{1.6}[][][0.8]{1.6}\psfrag{1.8}[][][0.8]{1.8}\psfrag{2.0}[][][0.8]{2.0}
\psfrag{2.2}[][][0.8]{2.2}
\psfrag{400}[][][0.8]{400}\psfrag{500}[][][0.8]{500}\psfrag{600}[][][0.8]{600}
\psfrag{700}[][][0.8]{700}\psfrag{800}[][][0.8]{800}\psfrag{1000}[][][0.8]{1000}
\psfrag{1200}[][][0.8]{1200} \psfrag{x}[][][1]{$x$
(nm)}\psfrag{Potential}[][][1]{potential energy (mK)}
\includegraphics[totalheight=6cm]{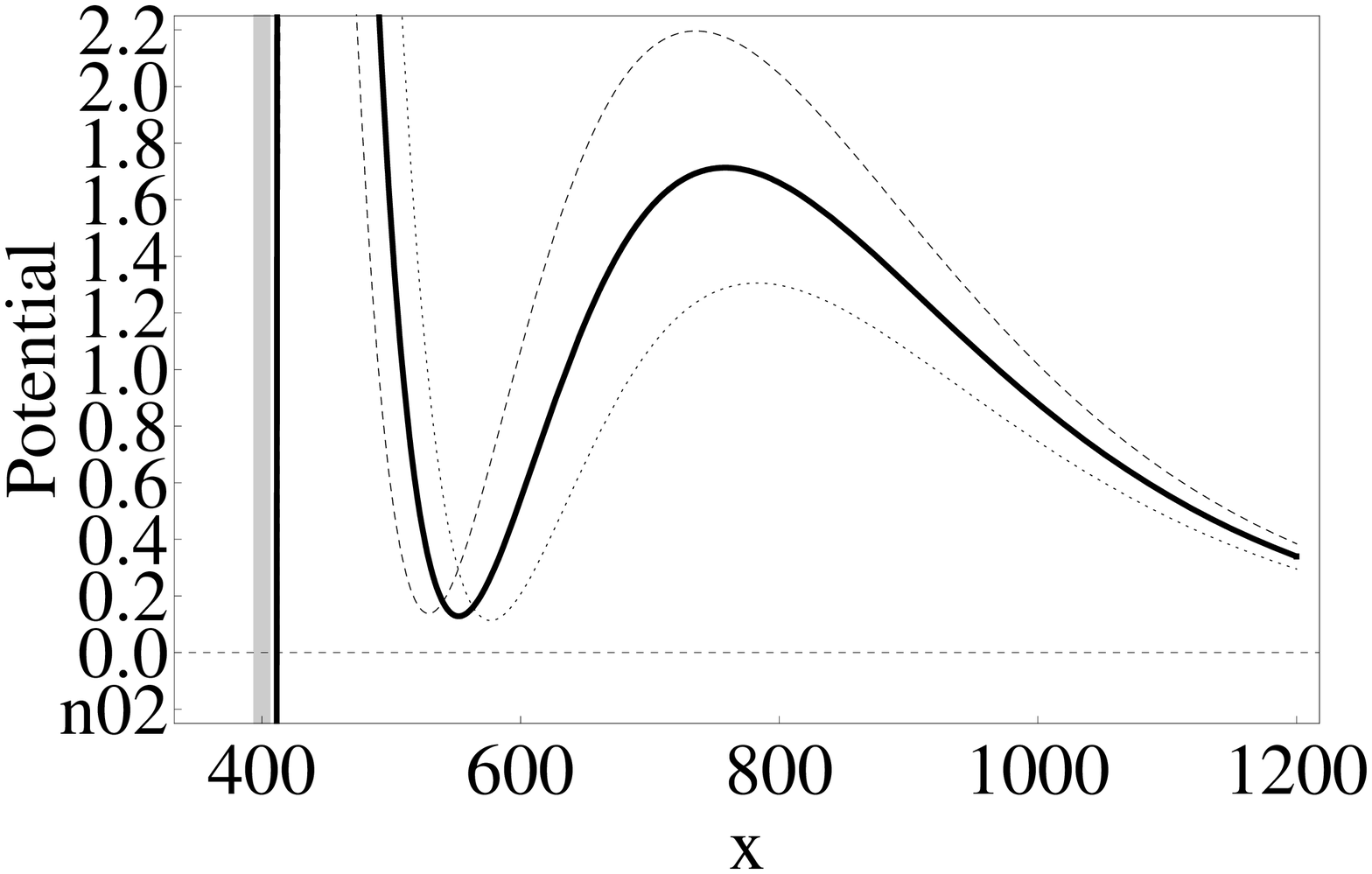}
\caption{Plot of the trapping potential versus the position along
the $x$ axis for P$_{11}=\tau_{0}$P (solid line),
P$_{11}=(\tau_{0}+\sigma)$P (dotted line) and
P$_{11}=(\tau_{0}-\sigma)$P (dashed line). The parameters are the
same as in Fig.~\ref{fig:HE21_HE11_az_all}. The fibre surface is
indicated by the vertical gray line.\label{fig:HE21_HE11_rad}}
\end{minipage}%
\begin{minipage}[t]{0.65\textwidth}
\centering\psfrag{n02}[][][0.8]{-0.2}\psfrag{0.0}[][][0.8]{0.0}\psfrag{0.2}[][][0.8]{0.2}
\psfrag{0.4}[][][0.8]{0.4}\psfrag{0.6}[][][0.8]{0.6}\psfrag{0.8}[][][0.8]{0.8}
\psfrag{1.0}[][][0.8]{1.0}\psfrag{1.2}[][][0.8]{1.2}\psfrag{1.4}[][][0.8]{1.4}
\psfrag{1.6}[][][0.8]{1.6}\psfrag{1.8}[][][0.8]{1.8}\psfrag{2.0}[][][0.8]{2.0}
\psfrag{2.2}[][][0.8]{2.2}
\psfrag{400}[][][0.8]{400}\psfrag{500}[][][0.8]{500}\psfrag{600}[][][0.8]{600}
\psfrag{700}[][][0.8]{700}\psfrag{800}[][][0.8]{800}\psfrag{n200}[][][0.8]{-200}
\psfrag{0}[][][0.8]{0}\psfrag{200}[][][0.8]{200}
\psfrag{x}[][][1]{$l$ (nm)}\psfrag{Potential}[][][1]{potential
energy (mK)}
\includegraphics[totalheight=6cm]{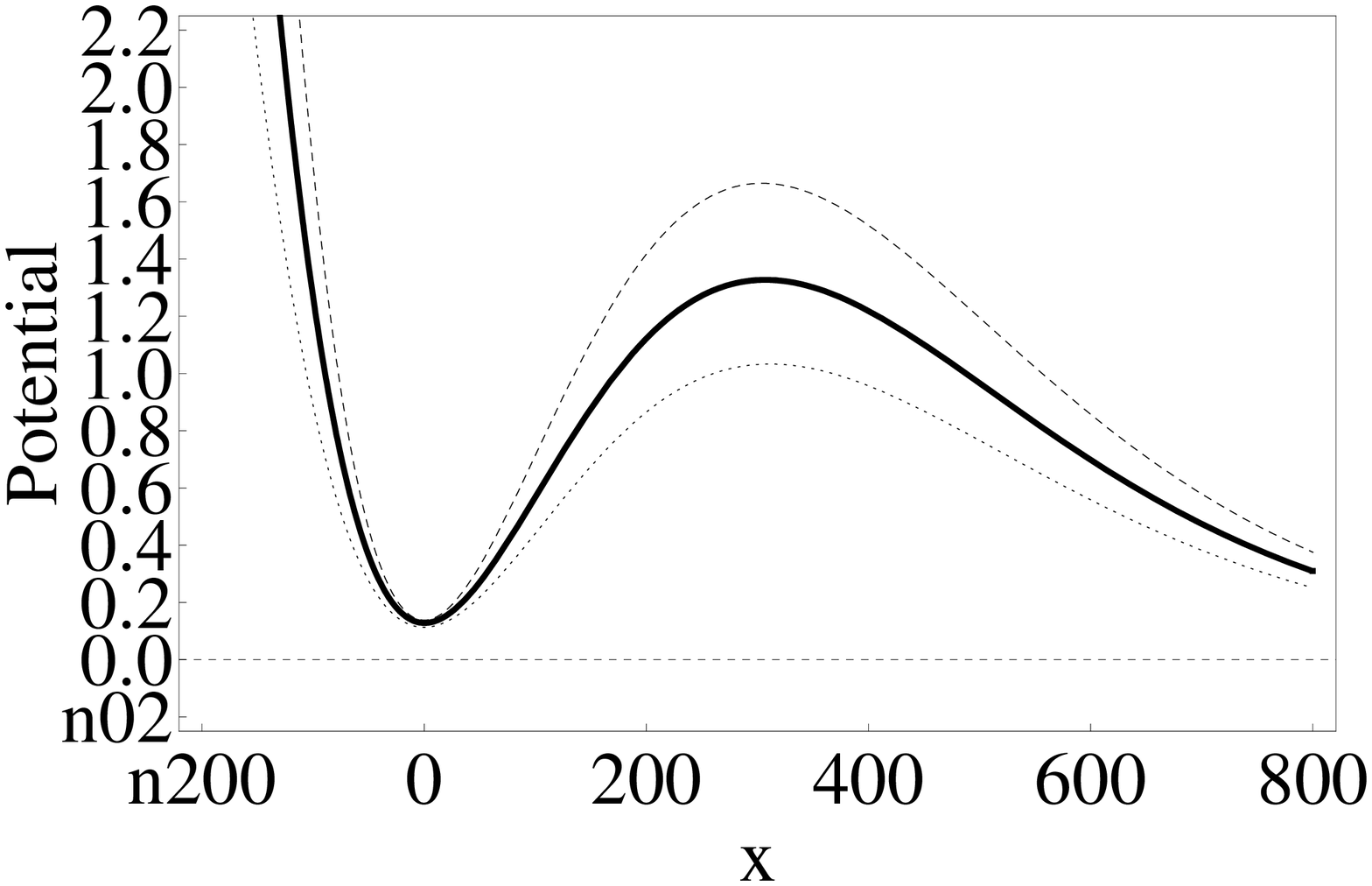}
\caption{Plot of the trapping potential versus the position along
the direction of minimal potential barrier $l(\tau)$ for
P$_{11}=\tau_{0}$P (solid line), P$_{11}=(\tau_{0}+\sigma)$P
(dotted line) and P$_{11}=(\tau_{0}-\sigma)$P (dashed line). The
parameters are the same as in
Fig.~\ref{fig:HE21_HE11_az_all}.\label{fig:HE21_HE11_rad_askew}}
\end{minipage}
\end{narrow}

\end{figure}

Figure~\ref{fig:HE21_HE11_az_all} shows a contour plot of the
trapping potential in the plane $z=3.45$~$\mu$m. The two dashed
lines with their origin at the center of the trap indicate the two
directions with minimal potential barrier which, by consequence,
determine the depth of the trap (see
Fig.~\ref{fig:HE21_HE11_rad_askew}). The trapping minimum is at
$\phi=0$, $r=552$~nm and $z=3.45$~$\mu$m. It lies on the $x$-axis
because here the polarisation of the two modes matches and the
interference is maximally destructive. This polarisation matching
between the two modes becomes apparent when comparing
Figs.~\ref{fig:Field_plot_HE11_lin}
and~\ref{fig:Field_plot_HE21_linear}. However, unlike the
HE$_{11}$+TE$_{01}$ trap considered in Sect.~\ref{HE11+TE01 trap},
the polarisation matching between the two modes is not perfect.
This is due to the fact that the ratio
$E_{z}/\arrowvert\vec{E_{\perp}\arrowvert}$ at the trapping
minimum is different for the two modes and, therefore, the
electric fields never cancel completely. Indeed, this stems from
the orientation of $\vec{E_{\perp}}$ at the position of the trap:
When the transverse electric field is perpendicular to the fibre
surface, a non-vanishing $z$-component of the electric field
arises~\cite{Snyder}. This polarisation configuration results in a
more intense evanescent field allowing the creation of a trap
comparable to the one presented in Sect.~\ref{HE11+TE01 trap} with
only 50\% of the power. As a drawback, the intensity at the
trapping minimum is not zero. When varying the parameters
$\varphi_{0}$ and $\phi_{0}$ in Eqs.~(\ref{eq:Exlinoutside})
to~(\ref{eq:Ezlinoutside}) and~(\ref{eq:HE21xlinoutside})
to~(\ref{eq:HE21zlinoutside}), respectively, the polarisation
direction of the two modes can be rotated and thereby the
azimuthal position of the trap can be varied. We calculate the
azimuthal oscillation frequency to be
$\omega_{\phi}/2\pi\approx330$~kHz. The extension of the trapping
volume in the azimuthal direction for caesium atoms with a
kinetic energy corresponding to 100~$\mu$K is 104~nm.\\

Figure~\ref{fig:HE21_HE11_long_all} shows the contour plot of the
trapping potential in the plane $y=0$. Like in the
HE$_{11}$+TE$_{01}$ trap, the interference between the modes
creates an axial array of traps with a periodicity given by the
beat length of the two co-propagating modes,
$z_{0}=2\pi/(\beta_{11}-\beta_{21})=3.45$~$\mu$m. Again, there is
a second array of traps at the opposite side of the fibre with
same periodicity and shifted by $z_{0}/2$. The potential has a
$\sin((\beta_{11}-\beta_{21})z)$ dependency in the axial direction
plus the offset due to the unbalanced $z$-components of the
electric fields of the two modes. We calculate the axial trapping
frequency to be $\omega_{z}/2\pi\approx610$~kHz. The extension of
the trapping volume in the axial direction for caesium atoms with
a kinetic energy
corresponding to 100~$\mu$K is 58~nm.\\

Figure~\ref{fig:HE21_HE11_rad} shows the trapping potential versus
the position along the $x$-axis. The solid black line shows the
radial trap for P$_{11}=\tau_{0}$P, the dashed line for
P$_{11}=(\tau_{0}-\sigma)$P, and the dotted line for
P$_{11}=(\tau_{0}+\sigma)$P, with $\tau_{0}=0.84$ and $\sigma=
0.018$. P denotes the total power propagating through the fibre
and P$_{11}$ the power propagating in the HE$_{11}$ mode. Again,
$\tau$ is assumed to be controlled with a precision of
$\sigma=0.05\sqrt{\tau_{0}(1-\tau_{0})}$. The light-induced
potential does not vanish at the minimum due to the mismatching in
the polarisation between the two modes. This leads to a higher
scattering rate of 57~photons/second compared to the trap
presented in Sect.~\ref{HE11+TE01 trap}. The radial trapping
frequency is $\omega_{r}/2\pi\approx970$~kHz. The extension of the
trapping volume in the radial direction for caesium atoms with a
kinetic energy corresponding to 100~$\mu$K is 37~nm. Note that the
depth of the potential shown in Fig.~\ref{fig:HE21_HE11_rad} does
not correspond to the depth of the trap because, as mentioned
above, the direction with minimal potential barrier for the atoms
is not radial. Figure~\ref{fig:HE21_HE11_rad_askew} therefore
shows the trapping potential against the position along the
direction with minimal potential barrier. The solid, dashed and
dotted lines have been calculated for the same values of $\tau$ as
in Fig.~\ref{fig:HE21_HE11_rad}. We define the direction with
minimal potential barrier $l$ as the straight line that connects
the potential minimum in the trap with the lowest local potential
maximum.  Note that $l$ depends on $\tau$ and has, per definition,
its origin at the trapping minimum. Hence, the three minima of the
potential profiles shown in Fig.~\ref{fig:HE21_HE11_rad_askew} are
located at $l=0$. The trap depth is then found to be $1.2$~mK. For
the case of P$_{11}=(\tau_{0}+\sigma)$P (dotted line) the trap is
33\% shallower compared to the trap for P$_{11}=\tau_{0}$P (solid
line), whereas for the case of P$_{11}=(\tau_{0}-\sigma)$P (dashed
line) the trap is 17\% deeper. Finally, we calculate a trapping
lifetime of 106 seconds for caesium atoms with an initial kinetic
energy corresponding to 100 $\mu$K. Again, the tunnelling through
the potential barrier in the radial direction towards the fibre
(see Fig.~\ref{fig:HE21_HE11_rad}) is negligible compared to the
lifetime of the atoms in the trap.\\

\section{HE$_{\textbf{{\scriptsize 21}}}$+TE$_{\textbf{{\scriptsize 01}}}$ trap}\label{HE21+TE01 trap}

Finally, we consider the trap arising from the interference
between the TE$_{01}$ and the HE$_{21}$ mode. It can be created
using 30~mW of light at a wavelength of 851.0~nm and the same
fibre parameters as in the above sections. The polarisation
orientation of the modes has been chosen such that the trap forms
at $\phi=3\pi/4$ and at $\phi=-\pi/4$. Note that this trapping
configuration has two trapping minima in the same $z$-plane,
whereas in the traps discussed before there is only one trapping
minimum per $z$-plane. The polarisation orientation corresponds to
$\phi_{0}=0$ in
Eqs.~(\ref{eq:HE21xlinoutside}),~(\ref{eq:HE21ylinoutside})
and~(\ref{eq:HE21zlinoutside}) for the HE$_{21}$ mode. With 68\%
of the power propagating in the TE$_{01}$ mode, i.e., $\tau=0.68$
(see Sect.~\ref{HE11+TE01 trap}), a trap for cold caesium atoms
with its trapping minimum at 184~nm from the fibre surface is
formed. The depth of the trap is 1.4~mK and, like in the above
cases, the trapping lifetime resulting from spontaneous scattering
of photons exceeds 100~seconds for caesium atoms
with an initial kinetic energy corresponding to 100~$\mu$K.\\

\begin{figure}[hbtp]
\begin{narrow}{-2cm}{2.5cm}
\begin{minipage}[t]{0.65\textwidth}
\centering\psfrag{x}[t][c][1]{$x$ (nm)} \psfrag{y}[b][c][1]{$y$
(nm)}
\psfrag{b1}[][][0.55]{0.6}\psfrag{b2}[c][b][0.55]{1.4}\psfrag{b3}[][][0.55]{10}
\psfrag{b4}[c][r][0.55]{30}\psfrag{b5}[][][0.55]{70}\psfrag{b6}[b][t][0.55]{1.4}
\psfrag{b7}[][][0.55]{0.6}
\psfrag{0}[][][0.8]{0}\psfrag{n500}[][][0.8]{-500}\psfrag{n1000}[][][0.8]{-1000}
\psfrag{500}[][][0.8]{500}\psfrag{1000}[][][0.8]{1000}
\includegraphics[totalheight=8.5cm]{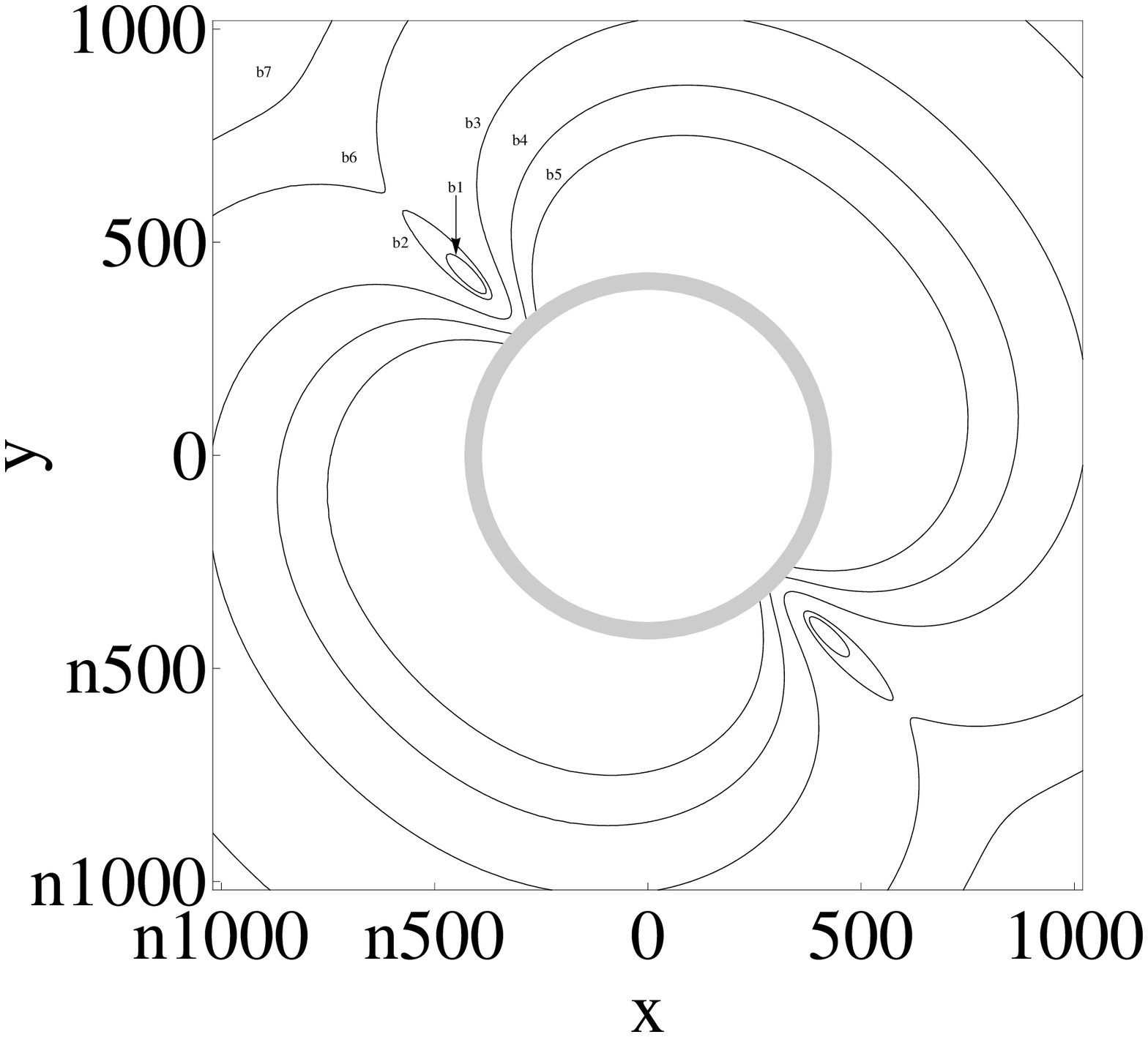}
\caption{Contour plot of the TE$_{01}$+HE$_{21}$ trap in the plane
$z=13.67$~$\mu$m for the following parameters: $P=30$~mW,
$\tau=0.68$, $\lambda=851.0$~nm, $a=400$~nm, $n_{1}=1.452$, and
$n_{2}=1$. The fibre surface is indicated by the gray circle and
the equipotential lines are labelled in
mK.\label{fig:TE01_HE21_az_all}}
\end{minipage}%
\begin{minipage}[t]{0.65\textwidth}
\centering\psfrag{x}[t][c][1]{$d$ ($\mu$m)}
\psfrag{z}[b][c][1]{$z$ ($\mu$m)}
\psfrag{b1}[c][c][0.55]{0.8}\psfrag{b3}[c][l][0.55]{1.4}
\psfrag{b4}[c][t][0.55]{4.5}\psfrag{b5}[b][rt][0.6]{20}\psfrag{b6}[c][c][0.55]{100}
\psfrag{b7}[c][t][0.55]{1.4}
\psfrag{0}[][][0.8]{0}\psfrag{n1}[][][0.8]{-1}\psfrag{nm5}[][][0.8]{-0.5}
\psfrag{0.5}[][][0.8]{0.5}\psfrag{1}[][][0.8]{1}
\psfrag{7}[][][0.8]{7}\psfrag{9}[][][0.8]{9}\psfrag{11}[][][0.8]{11}\psfrag{13}[][][0.8]{13}
\psfrag{15}[][][0.8]{15}\psfrag{17}[][][0.8]{17}\psfrag{19}[][][0.8]{19}
\psfrag{16}[][][0.8]{16}\psfrag{14}[][][0.8]{14}\psfrag{12}[][][0.8]{12}

\includegraphics[totalheight=8.5cm]{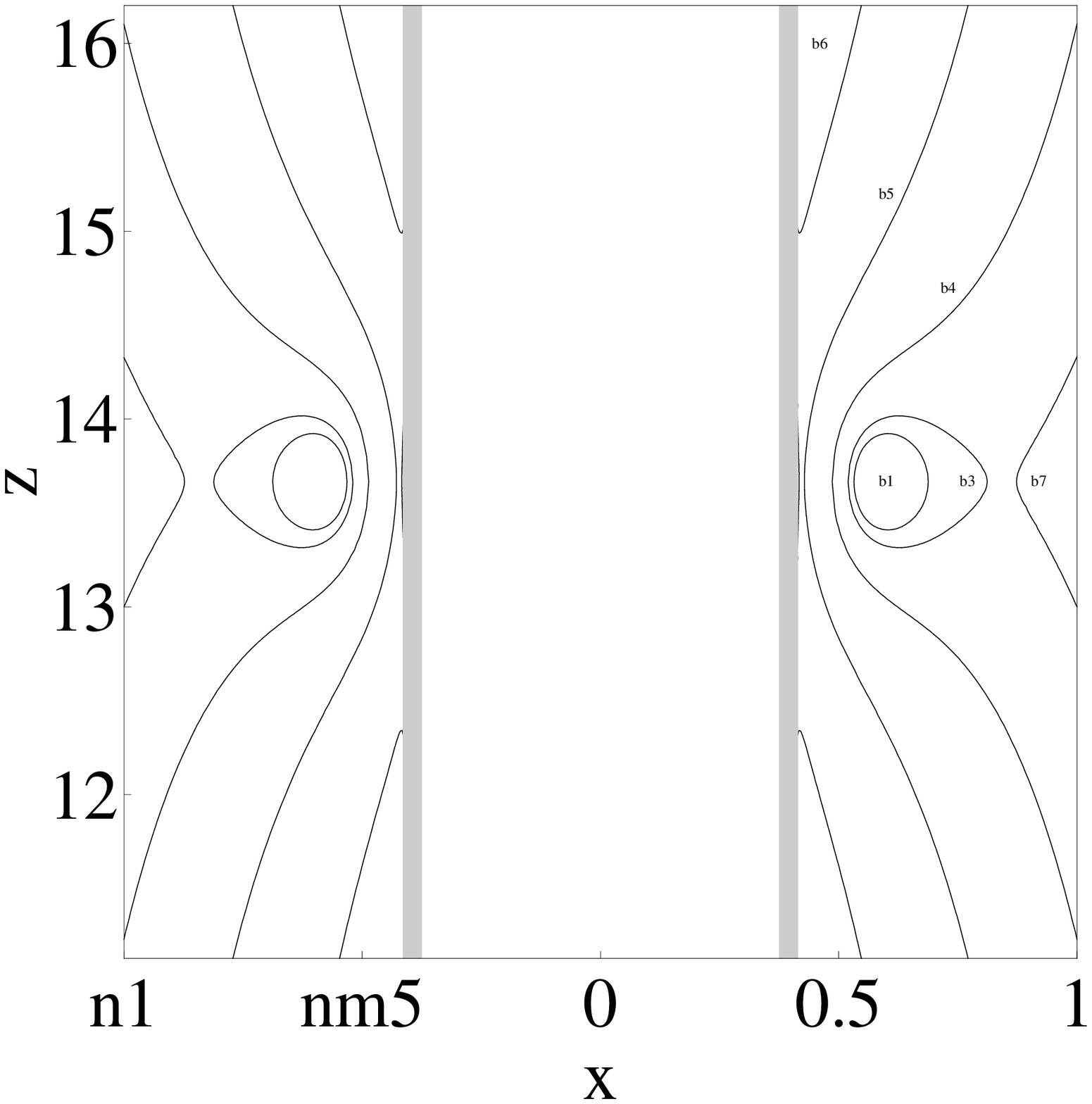}
\caption{Contour plot of the TE$_{01}$+HE$_{21}$ trap in the
$zd-$plane, where $d=(y-x)/\sqrt{2}$ for the same parameters as in
Fig.~\ref{fig:TE01_HE21_az_all}. The fibre surface is indicated by
the two vertical gray lines and the equipotential lines are
labelled in mK.\label{fig:TE01_HE21_long_all}}
\end{minipage}\\[20pt]
\end{narrow}

\centering\psfrag{n02}[][][0.8]{-0.2}\psfrag{0.0}[][][0.8]{0.0}\psfrag{0.2}[][][0.8]{0.2}
\psfrag{0.4}[][][0.8]{0.4}\psfrag{0.6}[][][0.8]{0.6}\psfrag{0.8}[][][0.8]{0.8}
\psfrag{1.0}[][][0.8]{1.0}\psfrag{1.2}[][][0.8]{1.2}\psfrag{1.4}[][][0.8]{1.4}
\psfrag{1.6}[][][0.8]{1.6}\psfrag{1.8}[][][0.8]{1.8}\psfrag{2.0}[][][0.8]{2.0}
\psfrag{2.2}[][][0.8]{2.2}
\psfrag{400}[][][0.8]{400}\psfrag{500}[][][0.8]{500}\psfrag{600}[][][0.8]{600}
\psfrag{700}[][][0.8]{700}\psfrag{800}[][][0.8]{800}\psfrag{1000}[][][0.8]{1000}
\psfrag{1200}[][][0.8]{1200} \psfrag{x}[][][1]{$d$
(nm)}\psfrag{Potential}[][][1]{potential energy (mK)}
\includegraphics[totalheight=8.5cm]{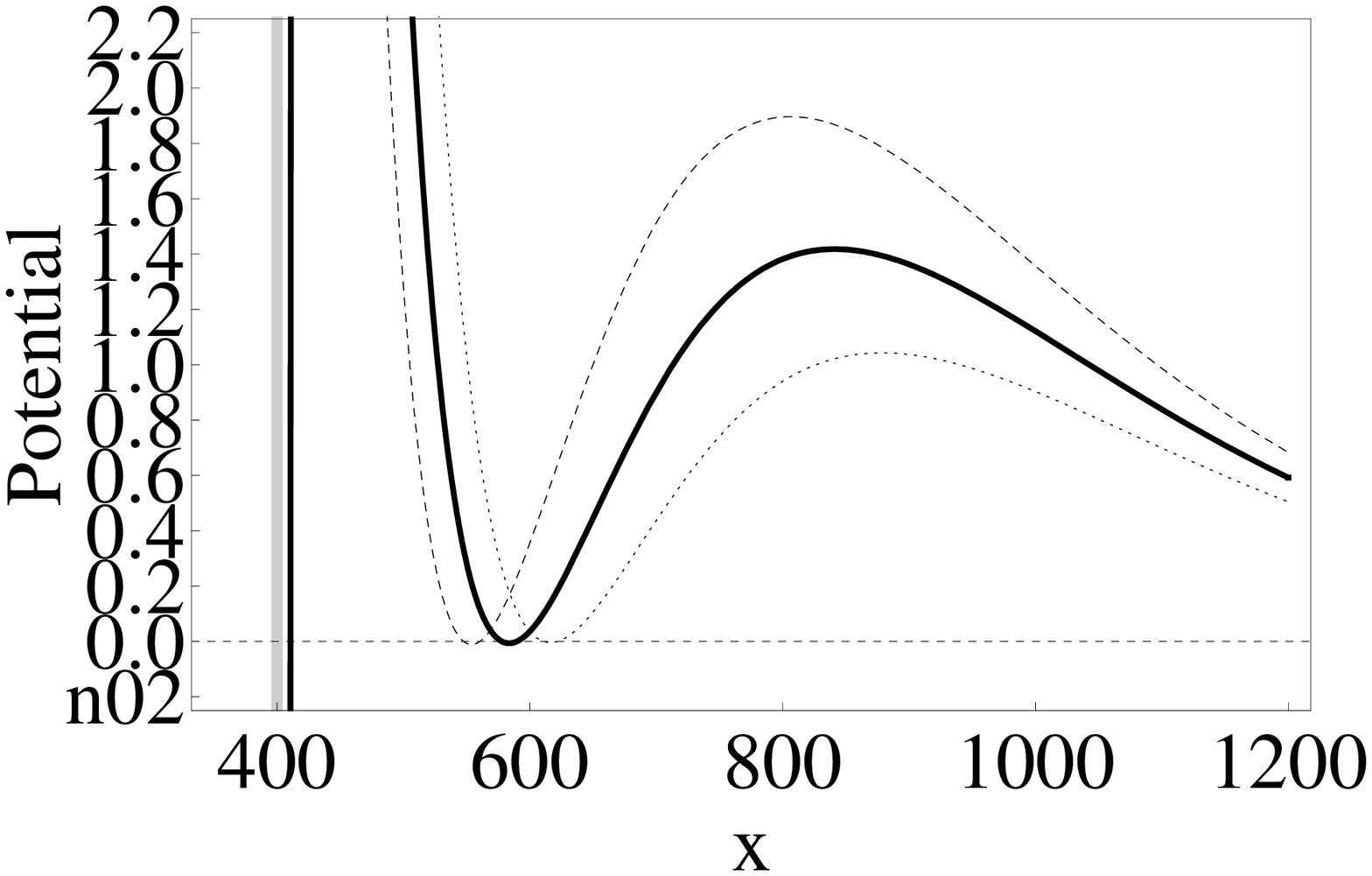}
\caption{Plot of the trapping potential versus the position along
the $d=(y-x)/\sqrt{2}$ axis for P$_{11}=\tau_{0}$P (solid line),
P$_{11}=(\tau_{0}+\sigma)$P (dotted line) and
P$_{11}=(\tau_{0}-\sigma)$P (dashed line). The parameters are the
same as in Fig.~\ref{fig:TE01_HE21_az_all}. The fibre surface is
indicated by the vertical gray line.\label{fig:TE01_HE21_rad}}

\end{figure}

Figure~\ref{fig:TE01_HE21_az_all} shows a contour plot of the trap
in the plane $z=13.67$~$\mu$m. Here, the trapping minima are shown
to be at $\phi=3\pi/4$, $r=584$~nm, $z=13.67$~$\mu$m and at
$\phi=-\pi/4$, $r=584$~nm, $z=13.67$~$\mu$m because the
polarisation in the two modes matches at these positions. This is
apparent when comparing Figs.~\ref{fig:Field_plot_TE01}
and~\ref{fig:Field_plot_HE21_linear}. Like for the
HE$_{11}$+TE$_{01}$ case, the polarisation matching between the
two modes is perfect because the $z$-component of the electric
field in the HE$_{21}$ mode vanishes at the position of the trap.
We calculate the azimuthal oscillation frequency to be
$\omega_{\phi}/2\pi\approx2.60$~MHz. The extension of the trapping
volume in the azimuthal direction for caesium atoms with a kinetic
energy corresponding to 100~$\mu$K is 14~nm. This strong
confinement in the azimuthal direction stems from the behaviour of
the polarisation of the electric field in the two the modes at the
position of the trap. When increasing $\phi$, the polarisation of
the HE$_{21}$ mode rotates clockwise, whereas the polarisation of
the TE$_{01}$ mode rotates anticlockwise. This produces a fast
polarisation mismatching between the two fields when displacing
the position along the azimuthal direction and thereby a steep
increase of the
potential.\\

Figure~\ref{fig:TE01_HE21_long_all} shows the contour plot of the
trap in the $zd-$plane, where $d=(y-x)/\sqrt{2}$. The interference
between the modes creates four axial arrays of traps with a
periodicity of $z_{0}=2\pi/(\beta_{01}-\beta_{21})=13.67$~$\mu$m.
The two trapping minima shown in Fig.~\ref{fig:TE01_HE21_az_all}
show the azimuthal positions of one pair of arrays. The second
pair is shifted with respect to the first one by $\phi=\pi/2$ and
$z=z_{0}/2$. We calculate the axial trapping frequency to be
$\omega_{z}/2\pi\approx204$~kHz. The extension of the trapping
volume in the axial direction for caesium atoms with a kinetic
energy corresponding to 100~$\mu$K is 174 nm. This elongation of
the trap compared to the traps presented in Sects.~\ref{HE11+TE01
trap} and~\ref{HE11+HE21 trap} stems from the large beat length
between the TE$_{01}$ and the
HE$_{21}$ mode.\\

Figure~\ref{fig:TE01_HE21_rad} shows the radial trapping potential
in the above defined $zd-$plane. The solid black line shows the
radial trap for P$_{01}=\tau_{0}$P, the dashed line for
P$_{01}=(\tau_{0}-\sigma)$P, and the dotted line for
P$_{01}=(\tau_{0}+\sigma)$P, with $\tau_{0}=0.68$ and
$\sigma=0.023$. Again, $\tau$ is assumed to be controlled with a
precision of $\sigma=0.05\sqrt{\tau_{0}(1-\tau_{0})}$. For the
case of P$_{11}=(\tau_{0}+\sigma)$P the trap is 25\% shallower
compared to the trap for P$_{11}=\tau_{0}$P, whereas for the case
of P$_{11}=(\tau_{0}-\sigma)$P the trap is 36\% deeper. Despite
the vanishing light-induced potential at the trapping minimum, the
total potential does not become significantly negative because the
influence of the van der Waals potential at this distance from the
fibre surface is negligible. We calculate the radial trapping
frequency to be $\omega_{r}/2\pi\approx770$~kHz. The extension of
the trapping volume in the radial direction for caesium atoms with
a kinetic energy corresponding to 100~$\mu$K is 47~nm. Since the
beat length between the TE$_{01}$ and the HE$_{21}$ mode is large
compared to the beat length in the other two traps, one would
expect the radial size of the trap to be large as well. However,
the difference in the decay lengths $\Lambda_{21}-\Lambda_{01}$ is
not the only factor that influences the radial profile of the
trap. It is also determined by the exact functional dependence of
the evanescent field for the different modes which results in a
similar radial confinement compared to the HE$_{11}$+TE$_{01}$ and
HE$_{11}$+HE$_{21}$ configurations. Finally, we calculate the
scattering rate and the trapping lifetime for caesium atoms with
an initial kinetic energy corresponding to 100~$\mu$K to be
62~photons/second and
114~seconds, respectively.\\

\section{Conclusions}\label{conclusions}

We presented three blue-detuned surface traps for cold atoms based
on two-mode interference between the lowest order modes in the
evanescent field around a 400-nm radius optical fibre. The
trapping potential confines the atoms in all three dimensions:
Radially, thanks to the tailored evanescent field, axially, thanks
to the difference in the phase velocity, and azimuthal, because
the different modes have different polarisation distributions
around the fibre axis. The three traps have a depth of the order
of 1~mK and a trapping lifetime that exceeds 100~seconds for
caesium atoms with an initial kinetic energy equivalent to
100~$\mu$K. Moreover, we have shown that the three traps are
robust against experimental uncertainties in the power
distribution between the modes. Such an array of optical
microtraps in the evanescent field surrounding an optical fibre in
combination with the highly efficient coupling of the atoms to the
fibre modes provides a very promising framework, e.g., for
experiments of storing and retrieving light with atomic ensembles.
Finally, the selective excitation of the fibre modes is
experimentally conceivable: One of the most promising methods is
the generation of Gauss-Laguerre modes in free space that match
the modes in the fibre~\cite{Maurer07}. The mapping between the
Gauss-Laguerre modes in free space and the modes in an ultra-thin
optical fibre is therefore currently under investigation in our
group.

\section{Acknowledgements}

We gratefully acknowledge financial support by the Volkswagen
Foundation (Lichtenberg Professorship) and the European Science
Foundation (EURYI Award).

\end{document}